# Improving accuracy of the numerical solution of Maxwell's equations by processing edge singularities of the electromagnetic field

Igor Semenikhin

*Valiev Institute of Physics and Technology of the Russian Academy of Sciences,*
*Nakhimovsky prosp. 34, 117218 Moscow, Russia*
*isemenihin@gmail.com*

**Abstract.** In this paper we present a methodology for increasing the accuracy and accelerating the convergence of numerical methods for solving Maxwell's equations in the frequency domain by taking into account the behavior of the electromagnetic field near the geometric edges of wedge-shaped structures. Several algorithms for incorporating treatment of singularities into methods for solving Maxwell's equations in two-dimensional structures by the examples of the analytical modal method and the spectral element method are discussed. In test calculations, for which we use diffraction gratings, the significant accuracy improvement and convergence acceleration were demonstrated. In the considered cases of spectral methods an enhancement of convergence from algebraic to exponential or close to exponential is observed. Diffraction efficiencies of the gratings, for which the conventional methods fail to converge due to the special values of permittivities, were calculated.

## 1. Introduction

It is well known that the electromagnetic field near geometric edges, both conductors and dielectrics, may possess singularities [1]. This leads to a slow convergence and sometimes even to the inability to obtain a solution in the seemingly simplest cases by conventional methods. For example, not so long ago the authors of [2] found that the calculation of the scattering of a TM polarized electromagnetic field on a lamellar diffraction grating (Fig. 1a) with certain values of the permittivities is not possible even by means of such proven methods as the Fourier Modal Method (FMM, RCWA) [3] and the Analytical Modal Method (AMM) [4,5,6].

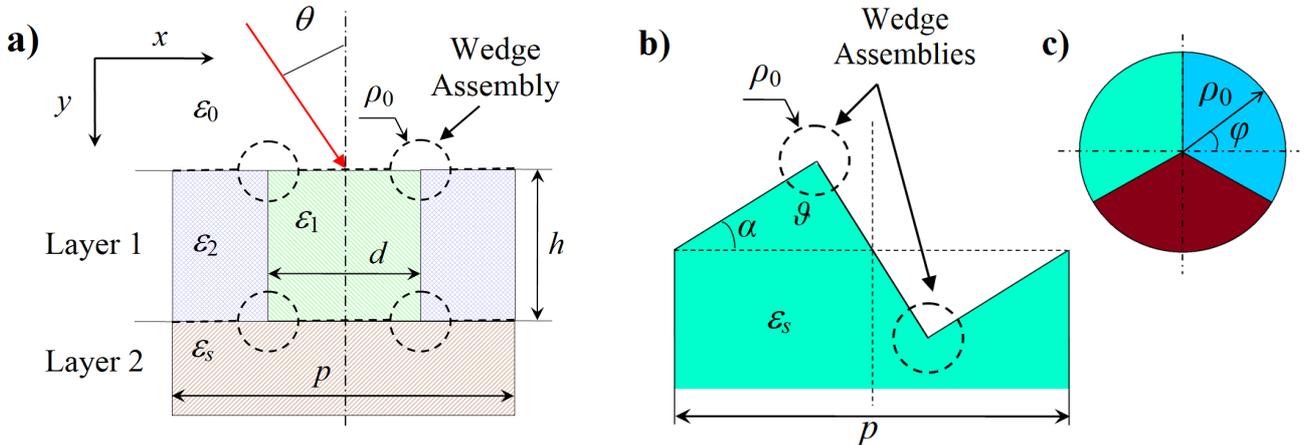

Fig. 1. Geometric parameters and permittivities of lamellar a) and triangular b) diffraction gratings. One period is depicted. The medium permittivity is $\varepsilon_0$, the incidence angle of the external electromagnetic field equals $\theta$. Dotted circles with a radius $\rho_0$ indicate the areas where the singularities arise (wedge assemblies); c) a wedge assembly in the general case.

It was shown in [7] that the same problem arises under certain conditions in the case of triangular diffraction gratings and generally in the case of any objects where external electromagnetic radiation is scattered at the intersection of faces with certain values of dielectric permittivity and with a certain value of the refractive index of the external environment. In [7,8] the reason of this was investigated and the conditions under which conventional calculation methods cease to converge were deduced. The singularities primarily occur in the components of the electromagnetic field lying in a plane perpendicular to the edge of wedge [1]. In particular, for TM polarization the components of the electric field near the edge of diffraction gratings asymptotically change proportionally to $\rho^{\tau_1-1}$, while the magnetic component parallel to the edge changes as $a + b(\varphi) \cdot \rho^{\tau_1}$, where $\rho$ is the distance to the edge in the polar coordinate system $(\rho, \varphi)$, the constants $a$, the function $b(\varphi)$ and the value of $\tau_1$ characterizing the singularity depend on the grating parameters and the refractive index of the external medium. In number of simple cases the value of $\tau_1$ can be obtained analytically [9,10], in the rest it is necessary to solve the transcendental equation numerically. In general, $\tau_1$ is a complex value and $\rho^{\tau_1-1} = \rho^{\text{Re}(\tau_1)-1} \exp[i \, \text{Im}(\tau_1) \ln \rho]$. For $0 < \text{Re}(\tau_1) < 1$ and a finite imaginary part of $\tau_1$ the oscillations of electric components with decreasing $\rho$ tend to infinity both in amplitude and frequency. In case of $\text{Re}(\tau_1) \to 0$ and $\rho \to 0$ simultaneously, the magnetic component also has no limiting value and oscillates with an increasing frequency around a certain constant. The closer $\text{Re}(\tau_1)$ is to zero, the slower the numerical methods converge, moreover, at $\text{Re}(\tau_1) = +0$ the conventional calculation methods such as FMM cease to work. Exactly such $\tau_1$ was considered in [2]. Hereinafter, by the term +0 we denote an infinitesimal positive value. Some modifications of FMM were proposed in [2,11] to solve the problem in the case of lamellar gratings. However, the authors of [8] expressed doubt about the reliability of the obtained results; besides, that did not solve the problem in the whole. The authors of [8] also noted that they do not know the methods currently in use that could give the correct result for $\text{Re}(\tau_1) = +0$. Of course, the case under consideration is specific, however, very often even in ordinary calculations it is necessary to obtain greater speed and accuracy than existing methods allow, which as a rule do not take into account edge singularities of the field explicitly.

To date, various approaches have been developed for processing edge singularities, including: Adaptive Spatial Resolution (ASR) in modal methods [12,13], mesh refinement toward the singularities and various adaptive strategies in the Finite Element Method [14,15], decomposition of the solution into a regular part and a singular part as in the Singular Field Method [16] and in the Singular Complement Method [17,18], the Weighted Regularization Method [19] and the method using of Perfectly Matched Layers at corners [20]. The simplest strategies, such as ASR, do not use additional information about the behavior of the field near the singularity. While more advanced ones, such as the Singular Complement Method, use the explicit knowledge of the expression of the singularities near the edge.

It must be said that the behavior of the electromagnetic field near the edge of wedge as in Fig. 1b) and near the edge of wedge assembly in Fig. 1a) and Fig. 1c) was studied quite a long time ago [21,22], since understanding the law of the field behavior, we can take it into account in the numerical calculation. For example, the convergence of spectral methods which are based on the expansion of the solution in smooth functions such as orthogonal polynomials or trigonometric functions, to which FMM and AMM belong, directly depends on the smoothness of the solution [23]. If in the region under consideration the solution and all its derivatives are continuous, a very fast, so-called exponential or spectral convergence is realized, when the higher expansion coefficients decrease exponentially. If the solution has singularities in the expansion domain, then the convergence will be, at best, algebraic, when the coefficients decrease according to a power law. If the solution in the region near the singularity is known, one can use the expansion in smooth functions outside this region, and then join these two solutions at the boundary, thus very fast convergence can be achieved.

In the original work [24] the field near the singularity was obtained in the form of a power series in $\rho$, so-called Meixner series, with coefficients calculated from a recurrent system of differential equations. It was later clarified [25] that in the case of angles that are multiples of $\pi/2$ this series also contains the logarithms of $\rho$, and that the logarithms appear in a more general case when angles are rational numbers multiple of $\pi$ [25,26].

Based on the results obtained earlier by other authors, we somewhat reformulate the method of calculating the field near the singularity for more convenient integration with solution methods used in the rest part of the domain. We implement the transition from the static case, where the problem re-duces to solving the eigenvalue problem similar to that arising in modal methods, to the case of finite wavelength. As examples of practical application the solution near the edges is added to several well-known methods. In the first case it is the AMM, where a system of linear equations for the coefficients of the transmitted and the reflected modes is modified as a result of taking into account the field singularity. In the second case it is the spectral element method, where the region containing the singularity acts as one of the elements. In addition, an algorithm for constructing the matrix of Dirichlet-to-Neumann map operator (DtN) is presented for the region containing the singularity. This allows adding singularity processing to any multi-domain method for solving Maxwell's equations in the two-dimensional case when the DtN algorithm can be applied to join computational domains [27]. In particular, we briefly touch on the boundary integral method [28] by which a verification of reference values is carried out for the examples presented in this work. In all cases considered, the singularity handling allows to significantly increase the accuracy and the computation speed, as well as achieve exponential or close to exponential convergence. Additionally, we performed calculations of the electromagnetic field in the cases $\mathrm{Re}(\tau_1) = +0$ analyzed in [7,8], for which it was not possible to obtain the result by conventional methods and the author knows only one method described in Ref. [20] which have allowed to tackle such problems.

## 2. Solution near the edge of a wedge assembly in polar coordinates

The general solution of Maxwell's equations for nonconical diffraction in the two-dimensional (2-D) case can be represented as a linear combination of two fundamental polarizations: the transverse electric polarization (TE) and the transverse magnetic (TM). In this paper, we shall focus on the case of TM polarization as on a more complex one; in the case of TE polarization, everything can be done in a similar way. For TM polarization only the $H_z$ component of magnetic field is not zero and the equation in the frequency domain for this component in polar coordinates is:

$$\rho^{-1}\partial_\rho\left(\rho\partial_\rho H_z\right) + \rho^{-2}\varepsilon(\varphi)\partial_\varphi\left[\varepsilon^{-1}(\varphi)\partial_\varphi H_z\right] + \varepsilon(\varphi)k_0^2 H_z = 0, \quad H_\rho = H_\varphi = 0, \text{ (TM)} \qquad (1)$$

where $\varepsilon(\varphi)$ is the relative dielectric constant, $k_0 = 2\pi/\lambda$, $\lambda$ denotes the wavelength of the incident radiation. Hereafter, for simplicity we set the relative magnetic permeability to unity. We shall solve Eq. (1) in a circle $\rho \leq \rho_0$, see Fig. 1). For convenience we introduce a new variable: $r = \rho/\rho_0$ and equation (1) takes the form:

$$r\partial_r\left(r\partial_r H_z\right) + \varepsilon(\varphi)\partial_\varphi\left[\varepsilon^{-1}(\varphi)\partial_\varphi H_z\right] + \varepsilon(\varphi)\gamma^2 r^2 H_z = 0, \qquad (2)$$

where $\gamma = 2\pi(\rho_0/\lambda)$ is the dimensionless constant much less than unity.

We shall seek the field $H_z$ in the form of a series:

$$H_z(r,\varphi) = \sum_{j=0}^{\infty} h_j G_j(r,\varphi), \qquad (3)$$

where the set of functions $\{G_j(r,\varphi)\}$ is the solutions of Eq. (2) for the boundary conditions that form the complete set of linearly independent functions (basis) $\{G_j(r=1,\varphi)\}$ on the boundary $r=1$ [1]. Then the coefficients $h_j$ can be found from the boundary conditions $b(\varphi) \equiv H_z(r=1,\varphi)$ of a particular problem: $b(\varphi) = \sum_{j=0}^{\infty} h_j G_j(r=1,\varphi)$. We shall use as such a basis a linear combination of angular eigenfunctions, which is obtained naturally when considering a static case.

When passing to the static case ($\lambda \to \infty$, $\gamma \to 0$), Eq. (2) for each function from $\{G_j(r,\varphi)\}$ turns into:

---

[1] Note that the system of functions $\{G_j(r,\varphi)\}$ itself is not a basis in a circle $r \leq 1$, since it can be used to represent only solutions of equation (2).

$$r\partial_r\left(r\partial_r G_j\right) = -\varepsilon(\varphi)\partial_\varphi\left[\varepsilon^{-1}(\varphi)\partial_\varphi G_j\right]. \tag{4}$$

It is solved by the separation of variables:

$$G_j(r,\varphi) = \Phi_j(\varphi) R_j^0(r), \tag{5}$$

where $\Phi_j(\varphi)$ are the eigenfunctions of the one-dimensional Helmholtz equation:

$$-\varepsilon(\varphi)\partial_\varphi\left[\varepsilon^{-1}(\varphi)\partial_\varphi \Phi_j(\varphi)\right] = \tau_j^2 \Phi_j(\varphi), \tag{6}$$

corresponding to the eigenvalues $\tau_j^2$. For definiteness, we shall chose the sign of $\tau_j$ so that $\operatorname{Re}(\tau_j) \geq 0$.

Equation (6) is completely analogous to the eigenvalue problem that arises in modal methods, for example, in case of a lamellar grating with replacing the $x$ coordinate by the azimuth angle $\varphi$ and the wavelength $\lambda \to \infty$ ($k_0 = 0$), see the next paragraph. The solutions of Eq. (6) in some simple cases can be found analytically (see Appendix 1). The main properties of the eigenfunctions $\Phi_j(\varphi)$, including their linear independence and completeness, which is important for representing through them an arbitrary condition on the boundary $r = 1$, as well as the orthogonality with weight $\varepsilon^{-1}(\varphi)$ to the adjoint eigenfunctions $\Phi_j^+(\varphi)$ corresponding to the differential operator adjoint to the operator in Eq. (6) are given in [6], and the general theory of such non-self-adjoint boundary eigenvalue problems is presented, for example, in [29].

Substituting Eq. (5) into Eq. (4) and taking into account Eq. (6), we obtain the equation for $R_j^0(r)$:

$$\left[r\partial_r(r\partial_r) - \tau_j^2\right] R_j^0(r) = 0, \tag{7}$$

with solution:

$$R_j^0(r) = a_j r^{\tau_j} + b_j r^{-\tau_j}. \tag{8}$$

For $\operatorname{Re}(\tau_j) > 0$ the term with a negative sign in exponent is rejected due to physical reason (the requirement of a finite energy in a finite volume)[1]. We set the coefficients $a_j = 1$ so that on the boundary $r = 1$ the functions $R_j^0(r)$ are equal to unity.

In case of $\gamma > 0$ we shall seek the functions $G_j(r,\varphi)$ in the form:

$$G_j(r,\varphi) = \sum_{k=0}^\infty \Phi_k(\varphi) R_{k,j}(r), \tag{9}$$

with the condition that upon passing to the static case $\gamma = 0$, there must be

$$\lim_{\gamma \to 0} R_{k,j}(r) = \delta_{k,j} R_j^0(r), \tag{10}$$

where $\delta_{k,j}$ is the Kronecker delta. Further, we find the functions $R_{k,j}(r)$ in the form of an absolutely convergent series that continuously depends on $\gamma$, when the wedge angles are the product of $\pi$ and rational numbers (see Appendix 2.2, the case of irrational multipliers is also briefly discussed there); therefore, for sufficiently small $\gamma$ the linear independence of $G_j(r=1,\varphi)$ Eq. (9) will be preserved. To get an equation for the functions $R_{k,j}(r)$, we substitute Eq. (9) into Eq. (2), taking into account Eq. (6):

$$\sum_{k=0}^\infty \left[r\partial_r(r\partial_r) - \tau_k^2 + r^2\gamma^2\varepsilon(\varphi)\right] \Phi_k(\varphi) R_{k,j}(r) = 0. \tag{11}$$

---

[1] This requirement can be written as: $\int_V \varepsilon|E|^2 + \mu|H|^2\, dv < \infty$, where $V$ is a finite cylindrical volume containing the edge, and the components of the electric field are $E_\rho \sim \rho^{-1}\partial_\varphi H_z$, $E_\varphi \sim \partial_\rho H_z$.

Left multiplying Eq. (11) by the complex conjugate adjoint eigenfunctions $\Phi_l^{+*}(\varphi)$ and integrating over $\varphi$ on the interval $[0, 2\pi]$ with the weight function $\varepsilon^{-1}(\varphi)$, we obtain

$$\left[ r\partial_r (r\partial_r) - \tau_k^2 \right] R_{k,j}(r) + r^2 \gamma^2 \sum_{l=0}^{\infty} c_{k,l} R_{l,j}(r) = 0, \qquad (12)$$

where matrix elements $c_{k,l}$ are

$$c_{k,l} = \int_0^{2\pi} \Phi_k^{+*}(\varphi) \Phi_l(\varphi) d\varphi, \qquad (13)$$

and we take into account the orthogonality of $\Phi_l^+(\varphi)$ and $\Phi_l(\varphi)$ with weight $\varepsilon^{-1}(\varphi)$ [5, 6].

The functions $R_{k,j}(r)$ can be found by the iteration method, taking the static case $R_j^0(r)$ as the zeroth-order approximation (see Appendix 2.1). The solution is obtained as a series in even powers of $\gamma$:

$$R_{k,j}(r) = R_j^0(r) \left[ \delta_{k,j} + \gamma^2 a_{k,j,1}(r) \cdot r^2 + \gamma^4 a_{k,j,2}(r) \cdot r^4 + \ldots \right], \qquad (14)$$

where $a_{k,j,i}(r)$ are either constants independent of $r$ or if $\tau_k = \tau_j + 2l$ holds, where $l$ is a positive integer, then $a_{k,j,i}(r)$, $i \geq l$ consist of the sum of the constants and the logarithms to an integer power less than the power of $r$. The second option is realized when the angles of the wedges are the product of $\pi$ and rational numbers. For small $r$, as well as for small $\gamma$ (large $\lambda$), $R_{k,j}(r)$ tends to the static case (10). In Appendix 2.2 we show that the series (14) absolutely converges for any finite values of $\gamma$ when the wedge angles are the product of $\pi$ and rational numbers, and we briefly discuss the case of angles that are multiples of an irrational factor of $\pi$; there are also formulas for practical calculations of the coefficients of the series, see Eq.(110)-(112). Since we know the behavior of the functions $R_{k,j}(r)$ for $r \to 0$, this functions can also be found by a direct solution of the Cauchy problem for a system of ordinary differential equations (Appendix 2.3).

Summarizing all the above, the solution near the edge of a wedge assembly inside the circle $\rho_0$ can be written as follows:

$$\begin{aligned} H_z(\rho, \varphi) &= \sum_{j=0}^{\infty} h_j G_j(\rho/\rho_0, \varphi) = \sum_{j=0}^{\infty} h_j \sum_{k=0}^{\infty} \Phi_k(\varphi) R_{k,j}(\rho/\rho_0) = \\ &= \sum_{j,k=0}^{\infty} h_j \Phi_k(\varphi) \cdot (\rho/\rho_0)^{\tau_j} \left[ \delta_{k,j} + a_{k,j,1}(\rho/\rho_0) \cdot (k_0 \rho)^2 + a_{k,j,2}(\rho/\rho_0) \cdot (k_0 \rho)^4 + \ldots \right] \end{aligned}, \qquad (15)$$

where $\tau_j$ are the square roots of the eigenvalues such that $\mathrm{Re}(\tau_j) \geq 0$, and $\Phi(\varphi)$ are corresponding eigenfunctions of the problem (6) arising in a static case ($\lambda = \infty$), $R_{k,j}(\rho/\rho_0)$ are the solutions of the system of ordinary differential equations (12), and the coefficients $h_j$ are found from the boundary conditions at $\rho = \rho_0$.

## 3. Incorporating the solution near the edge of a wedge assembly into the Analytical Modal Method

In this section we shall apply the above-described technique for calculating the electromagnetic field near edges to accelerate the convergence of the Analytical Modal Method which is the base representative of the class of modal methods. Within these methods the simulation domain, for example, a diffraction grating, is divided into parallel layers, in each of which the permittivity changes only in the plane of the layer and is assumed to be constant along the perpendicular direction allowing the separation of spatial variables, see Fig.1a). Within each layer the eigenmodes of the electromagnetic field are calculated and the general solution is then expressed by means of an eigenmode expansion. The expansion coefficients can be found by applying the proper boundary conditions. In the case of TM polariza-tion the magnetic field within each $l$-th layer ($l=0..L$, where $l=0$ corresponds to the superstrate layer and $l=L$ to the substrate) with relative permittivity $\varepsilon_l(x)$ satisfies the following equation:

$$\partial_y^2 H_z = -\varepsilon_l(x)\partial_x\left[\varepsilon_l^{-1}(x)\partial_x H_z\right] - k_0^2 \varepsilon_l(x) H_z, \quad H_x = H_y = 0. \quad (16)$$

Along $x$ axis the quasiperiodic Floquet boundary conditions are applied:

$$H_z(x+p, y) = \exp(i k_0 n_0 p \sin\theta) H_y(x, y), \quad (17)$$

where $p$ is the period of grating, $n_0$ is the refractive index of the superstrate and $\theta$ is the incidence angle of external radiation in the form of a plane wave (Fig. 1):

$$H_z^{inc} = \exp[i k_0 n_0 (x \sin\theta + y \cos\theta)]. \quad (18)$$

After separating the variables in Eq. (16), the solution within each $l$-th layer can be written as [4,6]:

$$H_z(x, y) = \sum_n \left[c_{l,n}^+ \exp(i k_0 \kappa_{l,n} y) + c_{l,n}^- \exp(-i k_0 \kappa_{l,n} y)\right] \psi_{l,n}(x), \quad (19)$$

where $c_{l,n}^+$ and $c_{l,n}^-$ are the amplitudes of forward and backward propagating modes, $\kappa_{l,n}$ and $\psi_{l,n}(x)$ are the square roots of eigenvalues and the eigenfunctions of the one-dimensional Helmholtz equation with the quasiperiodic Floquet boundary conditions (17):

$$\varepsilon_l(x) k_0^{-1} \partial_x \left[\varepsilon_l^{-1}(x) k_0^{-1} \partial_x \psi(x)\right] + \varepsilon_l(x)\psi(x) = \kappa^2 \psi(x). \quad (20)$$

The sign of square root in (19) is chosen so that its imaginary part would be positive, whereas for real positive $\kappa_{l,n}^2$ the positive square root should be used. Thus, the directions of propagation and decaying of modes coincide. Various methods for solving Eq. (20) can be found in [4,6,30,31] and the same methods are applicable for solving Eq. (6).

In the edges regions bounded by circles of radius $\rho_0$ and marked in Fig. 1a) by a dashed line the solution is written in the form (15). In numerical calculations, we restrict the sum (15) over $j$ to the first $N_w$ terms. The number of terms in the sum over $k$ is chosen to be larger than $N_w$, so that with their further increase the final calculation results do not change within the required accuracy. For simplicity the numbers $N_w$ will be taken the same for all wedge assemblies $w$.

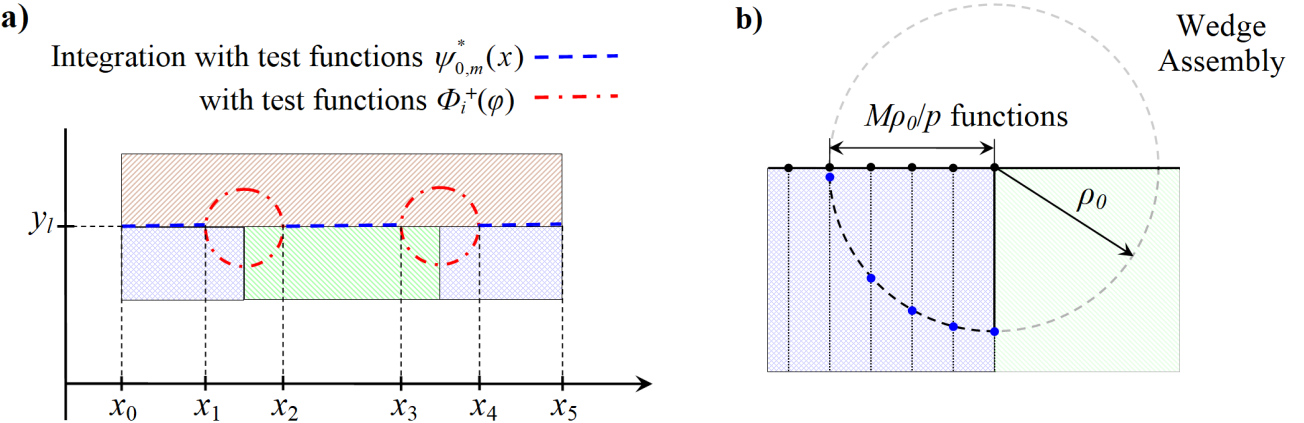

Fig. 2. a) Integration of conditions on the boundary (21) with test functions; b) the choice of radius $\rho_0$ for a given number of functions $N_w$ used to represent the solution near the edge of a wedge assembly.

The equations for the coefficients $c$ in Eq. (19) for each layer and the coefficients $h$ in Eq. (15) for each region near the edge are obtained by equating the values of the field and its normal derivative multiplied by $\varepsilon^{-1}$ on both sides of the interface $\Gamma$ indicated in Fig. 1a) by the dashed line:

$$H_z|_{\Gamma^+} = H_z|_{\Gamma^-}, \quad \varepsilon^{-1}\partial_\mathbf{n} H_z|_{\Gamma^+} = -\varepsilon^{-1}\partial_\mathbf{n} H_z|_{\Gamma^-}, \quad (21)$$

with a subsequent integration of these equations over a system of test functions. At the boundary of the layers on the segments $[x_{j-1}..x_j]$ between the edges Fig. 2a), indicated by the blue dashed line, we integrate Eq. (21) over the variable $x$ with the test functions for which we use the complex conjugate eigenfunctions of the superstrate $\psi_{0,m}^{+*}$ (Fourier matching scheme). At the boundary of the region around the edge, indicated in Fig. 2a) by the red dash-dotted line, we integrate Eq. (21) along the circle $\rho_0$ with the test functions $\Phi_i^{+*}(\varphi)$ and the weight $\varepsilon^{-1}(\varphi)$. More details on the preparation of equations for the coefficients can be found in Appendix 3.

The radius $\rho_0$ of the circle around the edge of a wedge assembly is chosen so that the ratio of the number of functions (9) to the length of the projection of the circle boundary onto the $x$ axis, see Fig. 2b), was approximately equal to the ratio of the number of modes to the period. The radius in this case is $\rho_0 = pN_w/(4M)$. At lower values of $\rho_0$ the convergence slows down; at large values, in addition, the condition number of the matrix of the resulting system of linear equations grows rapidly. The condition number increases for the same reason as in the Rayleigh method [32], due to the fact that exponentially decaying modes in Eq. (19), when integrated over the radius of the circle, give a very small contribution, which, for a finite length of the mantissa of numbers, is practically not distinguishable for different modes against the background of the remaining coefficients of the linear system.

| № | Relative permittivities | Geometric dimensions in units of wavelength $\lambda$ | $\theta$ | Tested characteristic and its reference value | References |
|---|---|---|---|---|---|
| 1 | $\varepsilon_0 = \varepsilon_2 = 1$, $\varepsilon_1 = \varepsilon_s = (0.22+6.71i)^2$ | $p = h = 1, d = p/2$ | $30^0$ | $R_0 = 0.8484816789046437$ | 31, 33, 34, 35 |
| 2 | $\varepsilon_0 = \varepsilon_2 = 1, \varepsilon_1 = 2.3^2, \varepsilon_s = 1.5^2$ | $p = 2, h = 1, d = 0.468$ | $30^0$ | $T_1 = 0.5105923632002064$ | 31, 34, 35, 36 |
| 3 | $\varepsilon_0 = 1.5^2, \varepsilon_1 = -2.5+0i$, $\varepsilon_2 = \varepsilon_s = 1.3^2$ | $p = 10/31, h = 14/31, d = p/2$ | $0^0$ | $R = 0.5942180852105747$ | 2,8,11 |

Table 1. Parameters of the lamellar diffraction gratings in Fig. 1a) used as examples. The TM-polarized electromagnetic wave is incident at an angle $\theta$.

As practical examples we calculated the diffraction efficiencies of several lamellar gratings in Fig. 1a). Table 1 shows the parameters of these gratings and references to some works in which they were previously used for such tests. In this paper we do not touch on hypersingularity [37] when $\text{Re}(\tau_1) = 0$ therefore, in instance, for the dielectric constant of grating No. 3 in table 1 we write $\varepsilon_1 = -2.5+0i$, which leads to the corresponding choice of the sign of the complex number $\tau_1$, such that $\text{Re}(\tau_1) = +0$.

Fig. 3 presents a comparison of the convergence of the modal methods often used in the calculation of such structures in the case of a metal diffraction grating No. 1 of Table 1. Relative error of the reflected zero-order diffraction efficiency $R_0$ calculated by the Fourier modal method (FMM), by the parametric Fourier modal method (PFMM) [12], and by the Analytical Modal Method (AMM), both with and without ($N_w = 0$) implementing expansion (15) near the edges, are shown. The value of the adaptive spatial resolution parameter $\eta = 0.995$ in the PFMM method is selected from the condition of best convergence. The value $R_0 = 0.8484816789046437$, calculated using multiple-precision arithmetic and rounded to 16 significant digits, was used as a reference. The reference values used in the work were verified by several methods, including the boundary integral method with incorporated edge singularity processing; for more detail, see the Section 4 of the article.

From Fig. 3 it can be seen that the FMM, PFMM methods and the regular version of AMM ($N_w = 0$) significantly lose in convergence rate to the modified AMM method, which explicitly takes into account edge singularities. In addition, in the case of a proportional increasing the number of functions $N_w$ in expansion (15) simultaneously with the number of modes $M$ the modified AMM method shows exponential or close to exponential convergence (Fig. 3a), while the rest of the presented spectral methods are limited by algebraic convergence due to the presence of edge singularities.

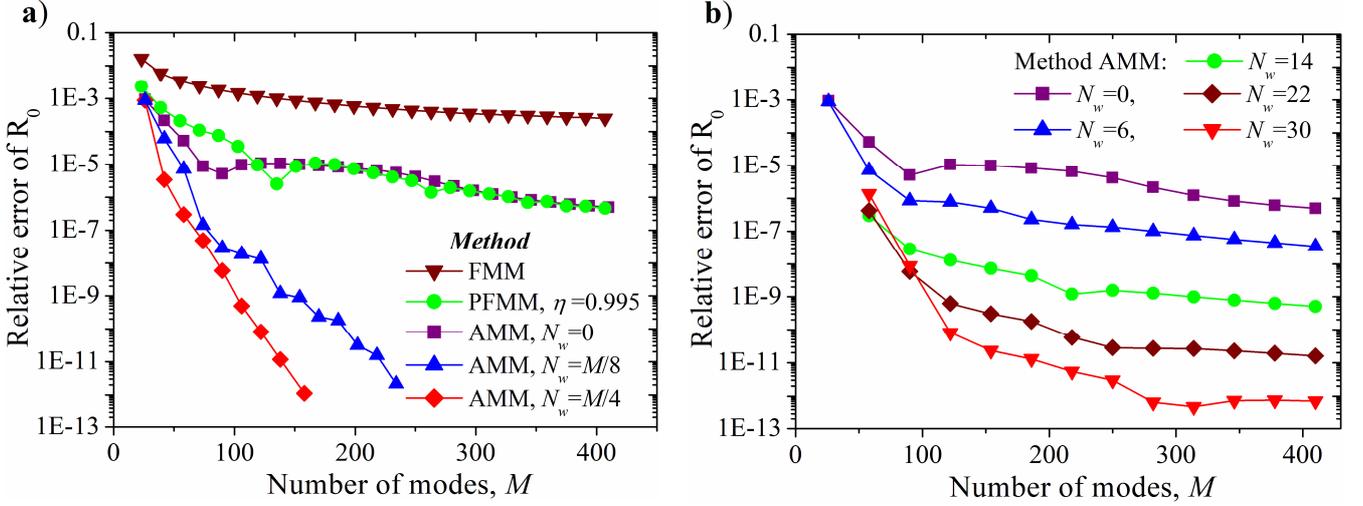

Fig. 3. a) Relative error of the reflected zero-order diffraction efficiency $R_0$ of the lamellar diffraction grating No. 1 of Table 1 versus the number of modes $M$ calculated by various modal methods; b) the results of calculation using the modified AMM method with a fixed number of functions $N_w$ used for representing the solution near the edges.

In the case of a fixed number of functions $N_w$ in the expansion (15) of the solution near the edges the convergence of the method is algebraic, Fig. 3b). Besides, with increasing $N_w$ the rate of convergence grows significantly. It can be noted that if earlier, in order to achieve relative accuracy $10^{-12}$ by the usual AMM method $M = 3.2 \times 10^5$ modes were required in this example [31], now, to obtain the same accuracy using the modified AMM method, only a few hundred modes are enough.

Fig. 4 shows the calculation of the first-order transmitted diffraction efficiency $T_1$ of a dielectric lamellar grating No. 2 of Table 1. As in the case of the metal diffraction grating the processing of edge singularities significantly accelerates convergence. And with a simultaneous proportional increase in the number of functions $N_w$ together with the number of modes $M$, an exponential or close to exponential convergence is also observed in Fig. 4a).

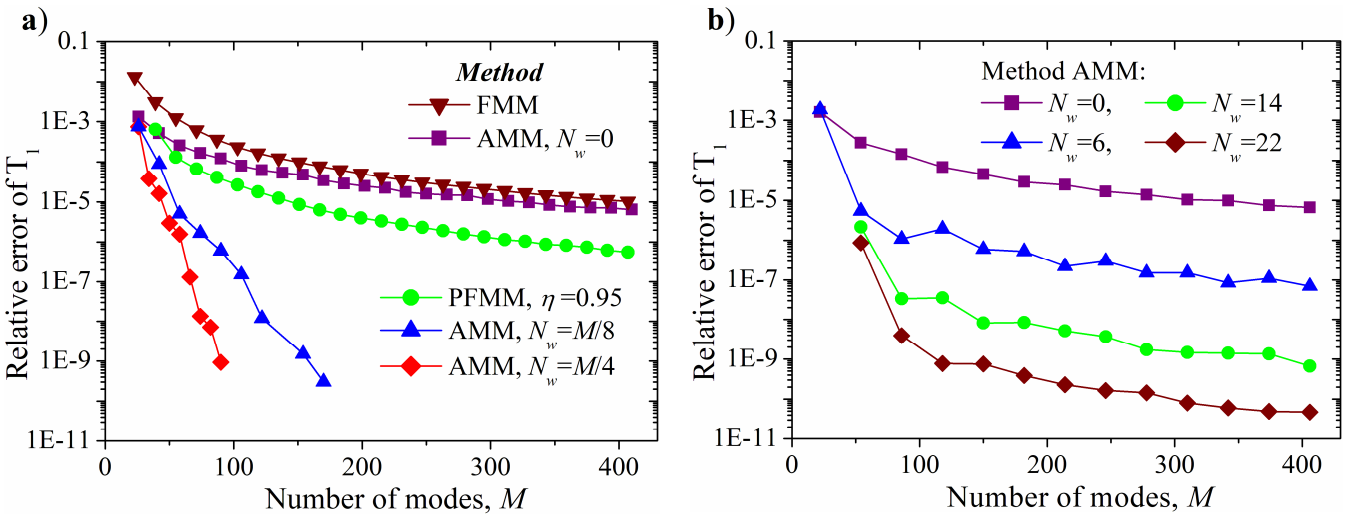

Fig. 4. a) Relative error of the first order transmitted diffraction efficiency $T_1$ of the dielectric lamellar diffraction grating No. 2 of Table 1 versus the number of modes $M$ calculated by various modal methods; b) the results of calculation using the modified AMM method with a fixed number of functions $N_w$ used for representing the solution near the edges.

In the following example, Fig. 5, the reflectivity of a lamellar diffraction grating No 3 of Table 1, featured the real part $\tau_1$ of all four edge singularities equal to +0, is calculated. It was previously studied in the papers [2,8,11], which we talked about in the introduction. Conventional methods such as FMM, PFMM, and AMM do not work in this case, and only the results of the modified AMM method are shown in the figures. If the singularity is explicitly taken into account, the value $\text{Re}(\tau_1) = +0$ no longer presents a problem, and the convergence of calculating the reflectance of the grating is as fast as in the previous examples.

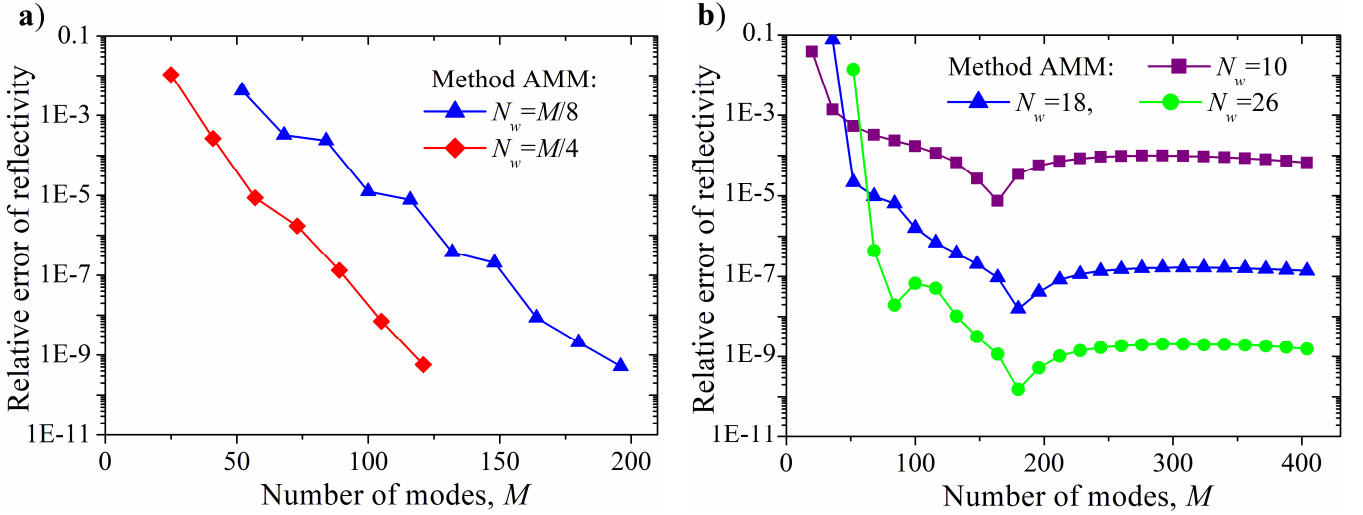

Рис. 5. a) Relative error of the reflectivity of the lamellar diffraction grating No. 3 of Table 1, where all edge singularities have $\text{Re}(\tau_1) = +0$, versus the number of modes $M$ calculated by the modified AMM; b) the same calculation but with a fixed number of functions $N_w$ used for representing the solution near the edges.

It should also be noted that the accuracy curves, Fig. 5b), calculated with a fixed number of functions $N_w$, tend practically to a constant with an increase in the number of modes $M$. This expresses the fact that in the case of $\text{Re}(\tau_1) = +0$ the achievable accuracy is limited not only by the number of used modes $M$, but also by the number of functions $N_w$ in the regions containing the edge.

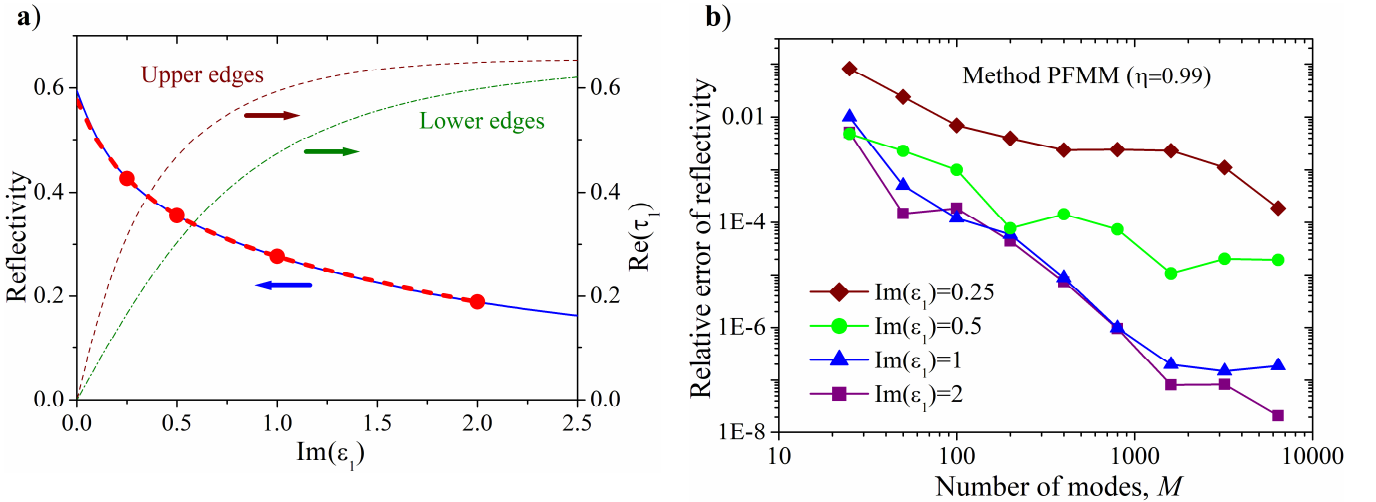

Fig. 6. a) Reflectivity (solid blue line), extrapolation by Thiele formula of reflectivity using four points (dashed red line) and the $\text{Re}(\tau_1)$ of the singularities at the upper edges (thin dashed line) and at the lower edges (thin dash-dotted line), see Fig 1a), of the diffraction grating No. 3 of table 1, depending on the imaginary part of the dielectric constant $\varepsilon_1$, while the real part $\varepsilon_1$ remains the same; b) the relative error in the calculation of reflectivity by the PFMM method for the four values of the imaginary part of the dielectric constant $\varepsilon_1$, which were used in the extrapolation procedure at the panel a).

Fig. 6a) shows how the reflectivity and the real part of $\tau_1$ vary with a decrease in the imaginary part of the dielectric constant $\varepsilon_1$ of the central part of the diffraction grating No. 3 of Table 1 from 2.5 to zero while the real part $\varepsilon_1$ remains the same. The curves of reflectivity and of $\text{Re}(\tau_1)$ are quite smooth, and their derivatives remain finite at $\text{Re}(\tau_1) = +0$.

From fig. 6b) a reducing convergence rate of the conventional PFMM method is seen, with a decrease of the imaginary part of $\varepsilon_1$ and a corresponding decrease of $\text{Re}(\tau_1)$. Due to the presence of the edge singularities, the convergence of the PFMM is algebraic and the slope of the convergence line is relaxed in logarithmic coordinates, while the number of modes required to obtain the result with the same accuracy increases exponentially. For $\text{Re}(\tau_1) = +0$ the PFMM method ceases to work, but using extrapolation we can approximately estimate the value of the reflectivity at the point $\text{Im}(\varepsilon_1) = 0$. Thiele extrapolation, using the four points marked in Fig. 6a) in circles, gives a rounded to two significant digits value of $R$=0.58, which differs from the exact one by 2.4%. It is clear that such an extrapolation can only be carried out *a posteriori*, knowing in advance about the sufficient smoothness of the curve in the direction of the extrapolated point. If the location of points is not close enough to $\text{Im}(\varepsilon_1) = 0$, then, as will be shown in the last example of the next section, such an extrapolation procedure can give a physically incorrect result.

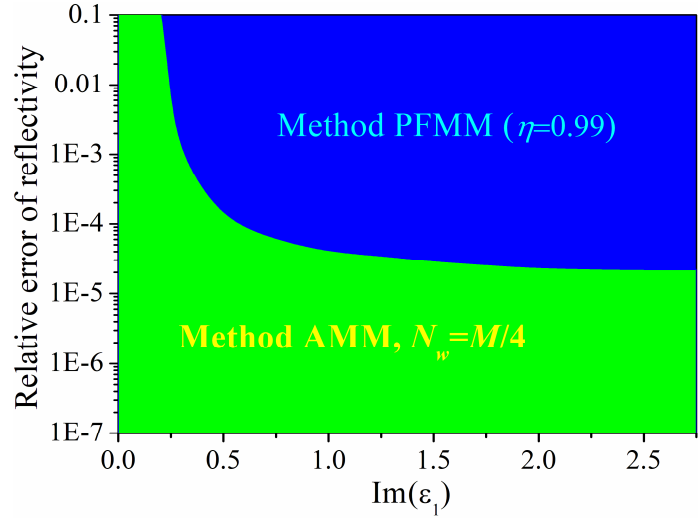

Fig. 7. The modified AMM method with the ratio $N_w = M/4$ between the number of functions $N_w$ and the number of modes $M$ is compared in terms of time consumption with the PFMM method with $\eta = 0.99$ in the case of diffraction grating No. 3 of Table 1, depending on the required calculation accuracy and the value of $\text{Im}(\varepsilon_1)$. In the area highlighted in blue it is more advantageous to use the PFMM method, in the green area the modified AMM method is preferable.

Until now, we have compared the convergence of the methods depending on the number of modes $M$. However, in real applications the time of calculation of the result with the required accuracy is primarily important. The calculation of the field near the corners in the form (15) takes a certain amount of time and, on the whole, the modified AMM method, which explicitly takes into account the edges singularities, works slower than the usual AMM method with the same $M$. The number of operations when calculating by formula (15) is proportional to $N_w^4$ and in the case when $N_w \sim M$ the total number of operations in the modified AMM method grows as $M^4$. While in most modal methods the number of operations is proportional to $M^3$, moreover, in the iterative implementation of the AMM method it is proportional to $M^2$ [31]. However, the convergence of the conventional methods is only algebraic, that is, the calculation errors is proportional to $M^{-q}$ with a certain constant $q$. At the same time, the convergence of the modified AMM method is exponential or close to exponential, and the calculation error is proportional to $e^{-c \cdot M}$ with some constant $c$. Therefore, with an increase in the result accuracy requirements, the modified method will ultimately win in terms of time consumption. Also, with decreasing of the values of $\text{Re}(\tau_1)$, conventional methods need an exponential increase in the number of modes to achieve the required accuracy, whereas the convergence of the modified method practically does not change. Thus, conventional methods, which do not take into account the edges singularities explicitly, can gain in computation time only if the required accuracy is moderate and the value of $\text{Re}(\tau_1)$ is not too small. As an example, we provide a diagram approximately showing when one method or another is preferable in the case of diffraction grating No. 3 of Table 1, depending on the required calculation accuracy and the value of $\text{Im}(\varepsilon_1)$. The modified AMM method is compared with the PFMM method in Fig. 7). This test was carried out using only one processor core without any parallelization.

## 4. Incorporating the solution near the edges into the Spectral Element Method

In this section, we shall incorporate the solution near the edges into the Spectral Element Method (SEM) [38], which is a universal method and which is very close to the finite element method. In this method the computational domain is divided into subdomains, in each of which the permittivity remains uniform or vary smoothly. The solution in the subdomains is represented in the form of an expansion over known basis functions, for which the orthogonal polynomials are usually used. In a subdomain containing an edge we shall represent the solution in the form of the series (15). The equations for the expansion coefficients in SEM are wrote in such a way that within each subdomain the solution satisfies Maxwell's equations, and at the boundaries with other subdomains it satisfies the corresponding conditions for the electromagnetic field. The coefficients can be calculated all at once by solving one large system of linear equations, which is usually done by some iterative method or, by implementing the Dirichlet-to-Neumann mapping (DtN) technique, where firstly the boundary conditions for the subdomains are found, and after that the Dirichlet problem is solved separately for each subdomain [27,39,40].

Speaking about the spectral element method, we adhere to the terminology of the book [23], referring to methods that use the spectral approach inside the subdomains, not only limiting by the Galerkin method in a weak form, as in the original proposal [38]. A formalism based on the application of the Dirichlet-to-Neumann (DtN) map, which we shall use, was previously discussed by us in [41], here we only briefly outline the main points.

In the case of TM polarization the magnetic field in each subdomain with uniform permittivity $\varepsilon$ satisfies the following equation:

$$\partial_x^2 H_z + \partial_y^2 H_z + \varepsilon k_0^2 H_z = 0, \quad H_x = H_y = 0. \qquad (22)$$

It can be solved both by the Galerkin method and by the pseudo-spectral method [23,39,40]. If a subdomain has not the rectangular form, then mapping into a rectangular area is applied by introducing new coordinates. For joining solutions in neighboring regions, it is convenient to use the Dirichlet-to-Neumann map [27,39,40]. To obtain the DtN map operator matrix for a subdomain, on the boundary $\Gamma$ of this subdomain a set of $N$ linearly independent boundary conditions at $N$ points of the Gauss or Gauss-Lobatto quadrature is defined and, by solving the Dirichlet problem for each of them, the corresponding normal derivatives at these points of the boundary are calculated. After that we compose two square matrices $N \times N$: in the first matrix $\mathbf{F}$ there will be $N$ boundary conditions as columns, and in the second matrix $\partial_\mathbf{n} \mathbf{F}$ there will be columns of the corresponding normal derivatives at the boundary. Since the boundary condition vectors are chosen linearly independent, an arbitrary boundary condition $\mathbf{f}_\Gamma$ defined at these points can be represented as their linear combination with some coefficients $\mathbf{c}$, and the corresponding normal derivative $\partial_\mathbf{n} \mathbf{f}_\Gamma$ can be easily calculated:

$$\mathbf{f}_\Gamma = \mathbf{F} \cdot \mathbf{c} \Rightarrow \partial_\mathbf{n} \mathbf{F} \cdot \mathbf{c} = \partial_\mathbf{n} \mathbf{f}_\Gamma \Rightarrow \partial_\mathbf{n} \mathbf{f}_\Gamma = \left( \partial_\mathbf{n} \mathbf{F} \cdot \mathbf{F}^{-1} \right) \mathbf{f}_\Gamma = \mathbf{D} \mathbf{f}_\Gamma, \qquad (23)$$

where **D** is the DtN map operator matrix. Boundary conditions, as well as normal derivatives, can be specified not only in the form of values at *N* points, but also in the form of *N* expansion coefficients for some fixed set of basis functions at the boundary, which is convenient if we are calculating a solution inside the subdomain using the Galerkin method. In this case, we shall say that DtN map acts in the space of these basis functions. The transition from one representation to another, as well as the calculation of the DtN map matrix for the subdomain containing the edge in various representations is given in Appendix 4.

Knowing the DtN map operator matrices $\mathbf{D}_1$ and $\mathbf{D}_2$ for two subdomains with a shared border $\Gamma'$:

$$\begin{pmatrix} \partial_\mathbf{n} \mathbf{f}_{1,\Gamma_1} \\ \partial_\mathbf{n} \mathbf{f}_{1,\Gamma'} \end{pmatrix} = \begin{pmatrix} (\mathbf{D}_1)_{1,1} & (\mathbf{D}_1)_{1,2} \\ (\mathbf{D}_1)_{2,1} & (\mathbf{D}_1)_{2,2} \end{pmatrix} \begin{pmatrix} \mathbf{f}_{1,\Gamma_1} \\ \mathbf{f}_{1,\Gamma'} \end{pmatrix}, \quad \begin{pmatrix} \partial_\mathbf{n} \mathbf{f}_{2,\Gamma_2} \\ \partial_\mathbf{n} \mathbf{f}_{2,\Gamma'} \end{pmatrix} = \begin{pmatrix} (\mathbf{D}_2)_{1,1} & (\mathbf{D}_2)_{1,2} \\ (\mathbf{D}_2)_{2,1} & (\mathbf{D}_2)_{2,2} \end{pmatrix} \begin{pmatrix} \mathbf{f}_{2,\Gamma_2} \\ \mathbf{f}_{2,\Gamma'} \end{pmatrix}, \tag{24}$$

and taking into account boundary conditions on $\Gamma'$, which for TM polarization are: $\mathbf{f}_{1,\Gamma'} = \mathbf{f}_{2,\Gamma'}$, $\partial_\mathbf{n} \mathbf{f}_{1,\Gamma'} = -\boldsymbol{\beta} \partial_\mathbf{n} \mathbf{f}_{2,\Gamma'}$, where the diagonal matrix $\boldsymbol{\beta} = \boldsymbol{\varepsilon}_1 / \boldsymbol{\varepsilon}_2$ is the ratio of the permittivities of the materials at points on both sides of the shared boundary $\Gamma'$, the DtN map operator matrix of these two elements combining can be straightforwardly calculated:

$$\begin{pmatrix} \partial_\mathbf{n} \mathbf{f}_{1,\Gamma_1} \\ \partial_\mathbf{n} \mathbf{f}_{2,\Gamma_2} \end{pmatrix} = \begin{pmatrix} (\mathbf{D}_1)_{1,1} + (\mathbf{D}_1)_{1,2} (\mathbf{D}'_1)_{2,1} & (\mathbf{D}_1)_{1,2} (\mathbf{D}'_2)_{2,1} \\ (\mathbf{D}_2)_{1,2} (\mathbf{D}'_1)_{2,1} & (\mathbf{D}_2)_{1,1} + (\mathbf{D}_2)_{1,2} (\mathbf{D}'_2)_{2,1} \end{pmatrix} \begin{pmatrix} \mathbf{f}_{1,\Gamma_1} \\ \mathbf{f}_{2,\Gamma_2} \end{pmatrix}, \tag{25}$$

where $(\mathbf{D}'_1)_{2,1} = \mathbf{C}(\mathbf{D}_1)_{2,1}$, $(\mathbf{D}'_2)_{2,1} = \mathbf{C}\boldsymbol{\beta}(\mathbf{D}_2)_{2,1}$, $\mathbf{C} = -\{(\mathbf{D}_1)_{2,2} + \boldsymbol{\beta}(\mathbf{D}_2)_{2,2}\}^{-1}$. Conversely, if the conditions on the boundaries $\Gamma_1, \Gamma_2$ are known, the conditions on the common boundary $\Gamma'$ will be:

$$\mathbf{f}_{1,\Gamma'} = \mathbf{f}_{2,\Gamma'} = (\mathbf{D}'_1)_{2,1} \mathbf{f}_{1,\Gamma_1} + (\mathbf{D}'_2)_{2,1} \mathbf{f}_{2,\Gamma_2}. \tag{26}$$

If we want to connect two adjacent boundaries $\Gamma_2$ and $\Gamma_3$ of the region characterized by a matrix of DtN map operator:

$$\begin{pmatrix} \partial_\mathbf{n} \mathbf{f}_{\Gamma_1} \\ \partial_\mathbf{n} \mathbf{f}_{\Gamma_2} \\ \partial_\mathbf{n} \mathbf{f}_{\Gamma_3} \end{pmatrix} = \begin{pmatrix} \mathbf{D}_{1,1} & \mathbf{D}_{1,2} & \mathbf{D}_{1,3} \\ \mathbf{D}_{2,1} & \mathbf{D}_{2,2} & \mathbf{D}_{2,3} \\ \mathbf{D}_{3,1} & \mathbf{D}_{3,2} & \mathbf{D}_{3,3} \end{pmatrix} \begin{pmatrix} \mathbf{f}_{\Gamma_1} \\ \mathbf{f}_{\Gamma_2} \\ \mathbf{f}_{\Gamma_3} \end{pmatrix}, \tag{27}$$

with conditions on the boundaries:

$$\mathbf{f}_{\Gamma_2} = \boldsymbol{\alpha} \mathbf{f}_{\Gamma_3}, \partial_\mathbf{n} \mathbf{f}_{\Gamma_2} = -\boldsymbol{\beta} \partial_\mathbf{n} \mathbf{f}_{\Gamma_3}, \tag{28}$$

where $\boldsymbol{\alpha}, \boldsymbol{\beta}$ are diagonal matrices as, for example, in the case of quasiperiodic boundary conditions (17), then, substituting Eq. (28) into Eq. (27), we obtain the DtN map for the remaining part of the boundary $\Gamma_1$:

$$\partial_\mathbf{n} \mathbf{f}_{\Gamma_1} = (\mathbf{D}_{1,1} + \mathbf{D}_{1,2} \boldsymbol{\alpha} \mathbf{D}' + \mathbf{D}_{1,3} \mathbf{D}') \mathbf{f}_{\Gamma_1},$$
$$\mathbf{D}' = -(\mathbf{D}_{2,2} \boldsymbol{\alpha} + \mathbf{D}_{2,3} + \boldsymbol{\beta} \mathbf{D}_{3,2} \boldsymbol{\alpha} + \boldsymbol{\beta} \mathbf{D}_{3,3})^{-1} (\mathbf{D}_{2,1} + \boldsymbol{\beta} \mathbf{D}_{3,1}). \tag{29}$$

And vice versa, after the conditions on the boundary $\Gamma_1$ are found, the conditions on the boundaries $\Gamma_2, \Gamma_3$ will be:

$$\mathbf{f}_{\Gamma_3} = \mathbf{D}' \mathbf{f}_{\Gamma_1}, \quad \mathbf{f}_{\Gamma_2} = \boldsymbol{\alpha} \mathbf{f}_{\Gamma_3}. \tag{30}$$

Thus, a DtN map operator matrix of the entire domain can be constructed from individual elements. After such a matrix is calculated, the unknown coefficients $\mathbf{c}_1^-$ of the reflected wave in superstrate and $\mathbf{c}_2^+$ of the transmitted wave in substrate can be found using expansion (19) of the field in the superstrate and in the substrate, and the definition of the DtN map at the upper and lower boundaries of the domain:

$$\begin{pmatrix} \mathbf{D}_{1,1} & \mathbf{D}_{1,2} \\ \mathbf{D}_{2,1} & \mathbf{D}_{2,2} \end{pmatrix} \begin{pmatrix} \mathbf{\Psi}_1 & 0 \\ 0 & \mathbf{\Psi}_2 \end{pmatrix} \begin{pmatrix} \mathbf{c}_1^+ + \mathbf{c}_1^- \\ \mathbf{c}_2^+ + \mathbf{c}_2^- \end{pmatrix} = ik_0 \begin{pmatrix} \mathbf{\Psi}_1 \mathbf{\kappa}_1 & 0 \\ 0 & -\mathbf{\Psi}_2 \mathbf{\kappa}_2 \end{pmatrix} \begin{pmatrix} -\mathbf{c}_1^+ + \mathbf{c}_1^- \\ -\mathbf{c}_2^+ + \mathbf{c}_2^- \end{pmatrix}, \qquad (31)$$

where the square matrices $\mathbf{\Psi}_l$ consist of the columns $\mathbf{\psi}_{l,j}$ of the functions $\psi_{l,j}(x_i)$ at the grid nodes. The coefficients $c_{1,j}^+ = \delta_{j,0}$, and $\mathbf{c}_2^- \equiv 0$ are known and specify the zero order incident wave in the superstrate and the upward going wave in the substrate, respectively. Instead of equating the field at grid points, the Galerkin method can be implemented by equating field integrals with test functions on both sides of the boundary, for more details see [41]. After the magnetic field at the boundary of the entire domain is obtained, we can, using Eq. (26) and Eq. (30), and acting in the opposite order, find the boundary conditions for each of the subdomains. Then, the field inside the subdomains can be calculated by solving the Dirichlet problem. In particular, the calculation of the coefficients $h_j$ in the expansion (15) of the magnetic field in the region near the edge is given in Appendix 4. It may be noticed that the DtN mapping approach is very similar to the **S**-matrix technique used in modal methods [42], with the difference that instead of reflection and transition coefficients, the field values at the subdomains boundaries are searched.

It should be noted that in the considered approach the DtN map matrices of individual subdomains can be calculated by using an arbitrary method, not necessarily by the spectral method. In the case of a uniform dielectric constant inside each element, it is very convenient to calculate the DtN map matrix by using the boundary integral method [28,36]. Since in this method the calculated subdomain does not need to be mapped into a rectangle, we get considerable freedom in choosing the partition into subdomains. In particular, we can represent regions of uniform permittivity as these subdomains. The subdomains containing the edges, as before, are calculated separately, and the whole technique of combining the subdomains remains the same. The boundary integral method scales much better with the number of collocation points increasing, and we employed its multiple-precision arithmetic implementation with incorporated edge singularity processing to verify the reference values used in the article.

As practical calculations shown, the value of the radius $\rho_0$ of the region containing the edge, where the solution is represented in the form of series (15), can be chosen in a first approximation so that the ratio of the number of functions $N_w$ to the length of the boundary of the region is equal to the average density of grid points per period, in accordance with the expression: $N_w / 2\pi\rho_0 = N_x / p$, where $N_x$ is the number of grid points along the $x$ axis over the grating period $p$. If the high accuracy result is required, then for faster convergence it is convenient to use the formula:

$$\rho_0 / p = c_1 N_w / N_x + c_2 (N_w / N_x)^2 + \ldots, \qquad (32)$$

where the coefficients $c_i$ can be approximately chosen from the condition of the best convergence of the method to the result of intermediate accuracy already obtained. In the work we used this formula with two coefficients $c_i$ in the series.

As examples, several triangular diffraction gratings Figs. 1b), the parameters of which are listed in Table 2, were used. Gratings with the same or similar characteristics were considered earlier in [7,8].

| № | Relative permittivities | Geometric characteristics | $\theta$ | Reference value |
|---|---|---|---|---|
| 1 | $\varepsilon_0 = 1, \varepsilon_s = -1.5 + 0i$ | $p = \lambda, \alpha = 30^0, \vartheta = 120^0$ | $0^0$ | $R_0 = 0.9983547969216146$ |
| 2 | $\varepsilon_0 = 1, \varepsilon_s = -4 + 0i$ | $p = \lambda/1.3, \alpha = 30^0, \vartheta = 60^0$ | $30^0$ | $R_{-1} = 0.7006512718026266$ |
| 3 | $\varepsilon_0 = 1, \varepsilon_s = -1.5 + 1.5i$ | $p = \lambda/0.9, \alpha = 30^0, \vartheta = 90^0$ | $60^0$ | $R_0 = 0.1161837518324407$ |

Table 2. Parameters of the triangular diffraction gratings Fig. 1b) used as examples. A TM-polarized electromagnetic wave is incident at an angle $\theta$.

Fig. 8a) shows the relative error of the zero-order reflected diffraction efficiency $R_0$ of a grating No. 1 of Table 2 versus the number of grid points $N_x$ along the x axis over the grating period p, calculated by the SEM with various numbers of functions $N_w$ in the subdomains containing edges. Previously this example was considered in [7], where the absence of the C-method (or Chandezon method [43]) convergence was observed due to the purely imaginary value of $\tau_1$. Note that in the case of a triangular grating Fig. 1b), the value of $\tau_1$ for the edge at the top vertex of the profile is the same as at the bottom; see the remark following the Eq. (41) in Appendix 1. It can be seen that the situation is the same as in the case of the analytical modal method: when the number of functions $N_w$ is constant, the convergence of the spectral method is algebraic, and is exponentially fast or close to it if $N_w$ is increasing proportionally with the number of nodes $N_x$.

In Fig. 8b) the spatial distribution of the squared magnitude of the magnetic field $|H(x,y)|^2$ is displayed. The color scale, shown on the right, is linear. The calculation was carried out at $N_x = 160$ grid points and at the number of functions $N_w = 18$. A thick white line marks the grating profile; thin lines indicate the boundaries of the subdomains. For better visibility of the boundaries of subdomains containing singularities, the radii $\rho_0$ of these subdomains in Fig. 8b) were taken an order of magnitude greater than the optimal ones, Eq. (32).

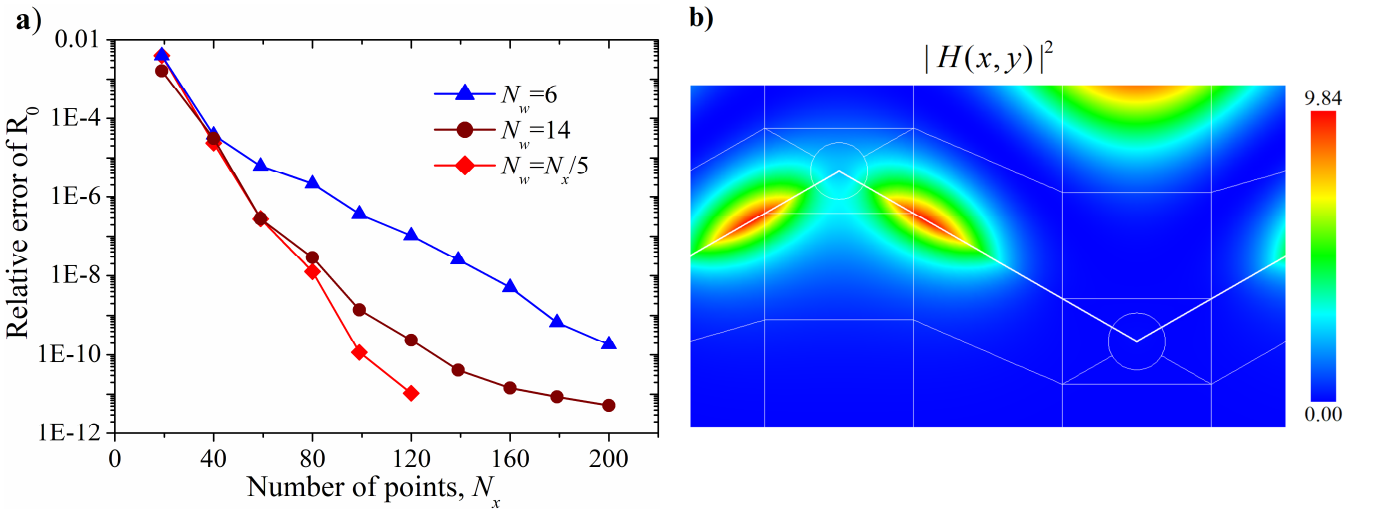

Fig. 8. a) Relative error of the reflected zero-order diffraction efficiency $R_0$ of a triangular diffraction grating No. 1 of Table 2 versus the number of grid points $N_x$ along the x axis over the grating period p, calculated by the spectral element method with various numbers of functions $N_w$ in the subdomains containing edges; b) the squared magnitude of the magnetic field $|H(x,y)|^2$. The thin white lines indicate the boundaries of the subdomains, and the thick one marks the grating profile.

It is seen that near the grating boundary, there are regions in which the magnetic field is much larger than the amplitude of the incident plane wave Eq. (18), which is equal to unity.

Fig. 9a) shows the relative error of the reflected minus first-order diffraction efficiency $R_{-1}$ of a triangular grating No. 2 of Table 2 versus the number of grid points $N_x$, calculated by the SEM with various numbers of functions $N_w$ in the subdomains containing edges. In this example, as in the previous one $\mathrm{Re}(\tau_1) = +0$, therefore, the conventional numerical methods also do not work here, which was pointed out in [7].

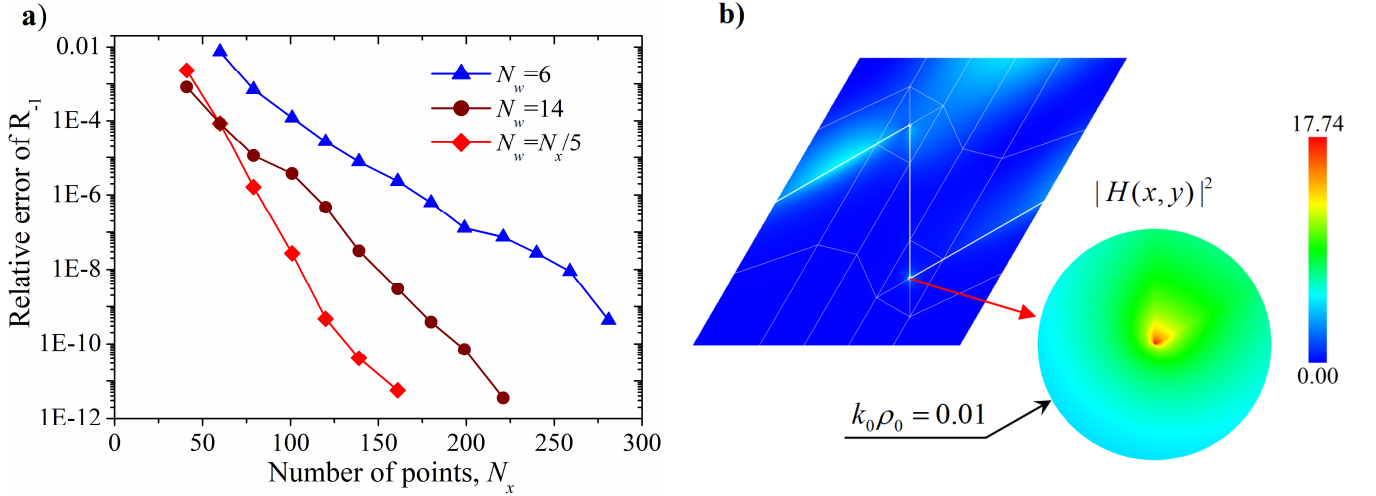

Fig. 9. a) Relative error of the reflected minus first-order diffraction efficiency $R_{-1}$ of a triangular diffraction grating No. 2 of Table 2 versus the number of grid points $N_x$ along the $x$ axis over the grating period $p$, calculated by the spectral element method with various numbers of functions $N_w$ in the subdomains containing edges; b) the squared magnitude of the magnetic field $|H(x,y)|^2$. The thin white lines indicate the boundaries of the subdomains, and the thick one marks the grating profile.

In Fig. 9b) the space distribution of the squared magnetic field is depicted. The optimal radii $\rho_0$, calculated by Eq. (32), of the subdomains containing singularities are very small, and the corresponding boundaries are not visible in the figure. The entire computational domain is inclined by an angle $[\alpha - (180^0 - \vartheta - \alpha)]/2$, which is equal to the difference in angles at the base of the triangle divided by 2, see Fig 1b). So that the angles of the subdomains are as close to $90^0$ as possible, this reduces the condition number of the linear system matrix when we solve the Dirichlet problem inside these subdomains. The magnitude of the magnetic field reaches its maximum near the lower angle of the grating, in the region containing the edge that is shown beside.

In Fig. 10a) the convergence of several methods was compared using the example of a triangular grating No. 3 of Table 2. We examined:

a) the conventional spectral element method without increasing the accuracy of the solution in the edge regions;

b) the parametric C-method with adaptive transformation of coordinates near edges (or Adaptive Spatial Resolution (ASR) as it is called in the original paper [13]) and a transformation parameter $\eta = 0.99$;

c) the spectral element method with adaptive transformation of coordinates (or adaptive spatial resolution) in subdomains near edges according to the law $\rho = \rho_0(1-\sin t)$, where the parameter $t$ changes from zero to $\pi/2$ at the edge point;

d) the spectral element method utilizing expansion (15) in the subdomains containing the edges and using the proportional increasing of the number of functions $N_w$ with the number of collocation points $N_x$.

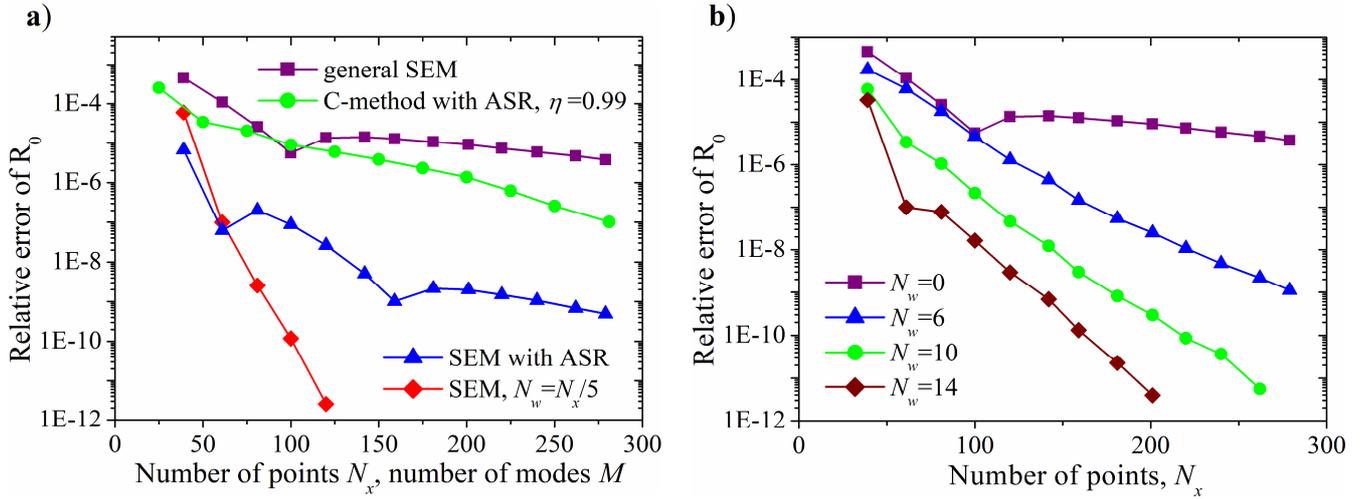

Fig. 10. a) Relative error of the reflected zero-order diffraction efficiency $R_0$ of a triangular diffraction grating No. 3 of Table 2 versus the number of grid points $N_x$ over the grating period $p$ for the spectral element method (SEM) and versus the number of modes $M$ for the parametric C-method with Adaptive Spatial Resolution (ASR); b) the results of calculation by the spectral element method with a fixed number of functions $N_w$ used for representing the solution near the edges.

The spectral element method employing expansion (15) demonstrates the best convergence. In Fig. 10b) the convergence of this method is shown for a different number of functions $N_w$ in the subdomains containing the edges.

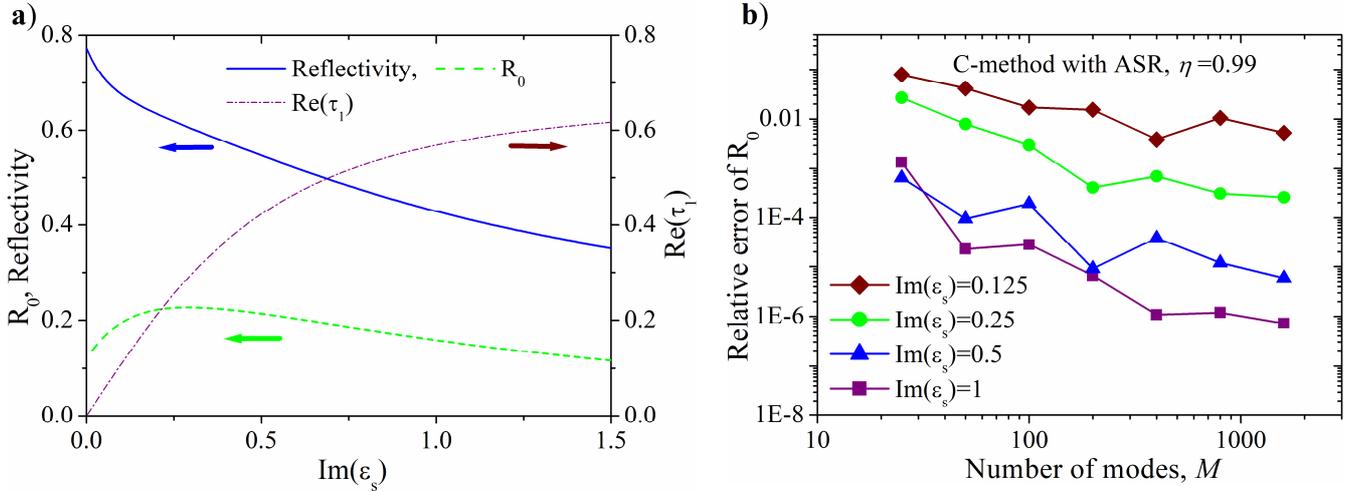

Fig. 11. a) Reflectivity, reflected zero-order diffraction efficiency $R_0$ and the real part of the $\tau_1$ versus the imaginary part of the dielectric constant $\varepsilon_s$ of a triangular diffraction grating No. 3 of Table 2, the real part $\varepsilon_s$ remains the same: $\mathrm{Re}(\varepsilon_s) = -1.5$; b) relative error of the reflected zero-order diffraction efficiency $R_0$ calculated by the parametric C-method for several values of the imaginary part of the dielectric constant $\varepsilon_s$.

Fig. 11 demonstrates how the reflected zero-order diffraction efficiency $R_0$, the reflectivity, and the real part of the $\tau_1$ change when the imaginary part of the dielectric constant $\varepsilon_s$ of grating No. 3 of Table 2 varies while the real part $\varepsilon_s$ remains the same. Just as in the last example of the previous section, the change in all examined quantities occurs smoothly and the derivatives stay finite at $\mathrm{Re}(\tau_1) = +0$. However, if we try to extrapolate $R_0$ by using the four values of $\varepsilon_s$, for which in Fig. 11b) graphs of convergence are presented, to $\mathrm{Im}(\varepsilon_s) = 0$ by utilizing the Thiele formula, similar to how it was done in the last example of the previous section, we get a physically incorrect result:

$R_0 < 0$. The polynomial extrapolation of $R_0$ using the same points gives a relative error of about 30%. To improve the result, it is necessary to shift the extrapolation points closer to $\text{Im}(\varepsilon_s) = 0$, which requires an exponential increase of the computation time in the case of conventional methods. Fig. 11b) demonstrates how the convergence rate of the parametric C-method diminishes, with a decrease in the imaginary part of $\varepsilon_s$ and a corresponding decrease in the real part of $\tau_1$.

For $\text{Re}(\tau_1) = +0$ the parametric C-method, like the conventional SEM, does not converge to a certain value. The reason for the divergence, consisting in the behavior of the field near the edges, has already been discussed in the papers [7,8]. Having the ability to calculate the field near the edges, we shall illustrate this using the example of the triangular grating No. 3 of table 2. In Fig. 12 the left panel shows the distribution of the squared magnitude of the magnetic field $|H(x,y)|^2$ for the dielectric constant of the grating material $\varepsilon_s = -1.5 + 1.5i$, and the right panel shows $|H(x,y)|^2$ for $\varepsilon_s = -1.5 + 0i$.

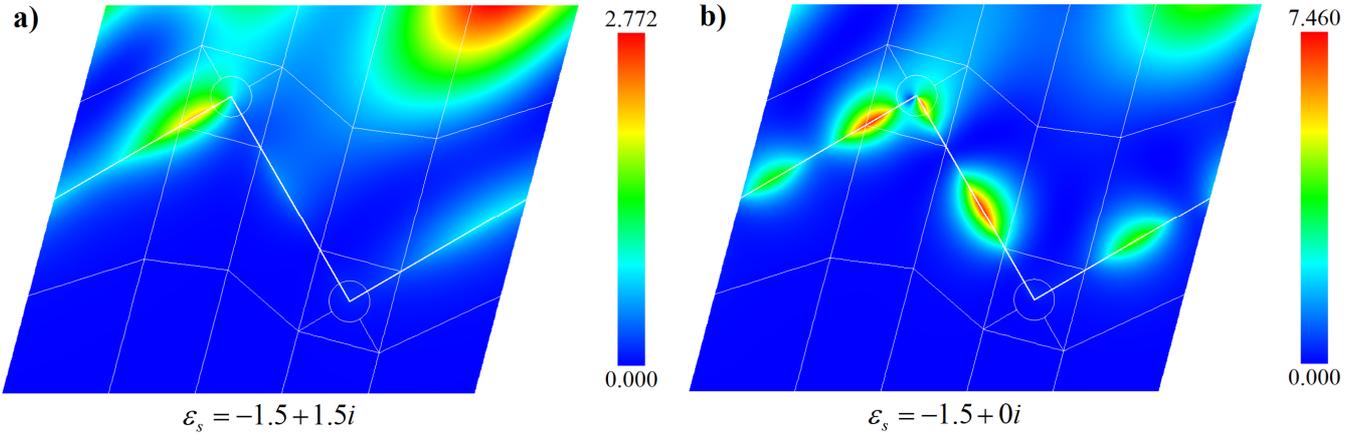

Fig. 12. The space distribution of the squared magnitude of the magnetic field $|H(x,y)|^2$ in the case of a diffraction grating No. 3 of Table 2 for two values of the dielectric constant of the grating material. The thin white lines indicate the boundaries of the subdomains, and the thick one marks the grating profile.

In the presented resolution the change in the field in both pictures seems quite smooth, and although in the second case its maximum value is several times larger, the behavior of the $|H(x,y)|^2$ distribution cannot say why the conventional numerical methods work in one case and not in the other. However, if we sequentially increase the scale of the region near the upper edge of Fig. 12b), we obtain a picture of the distribution of the squared magnitude of the magnetic field shown in Fig. 13.

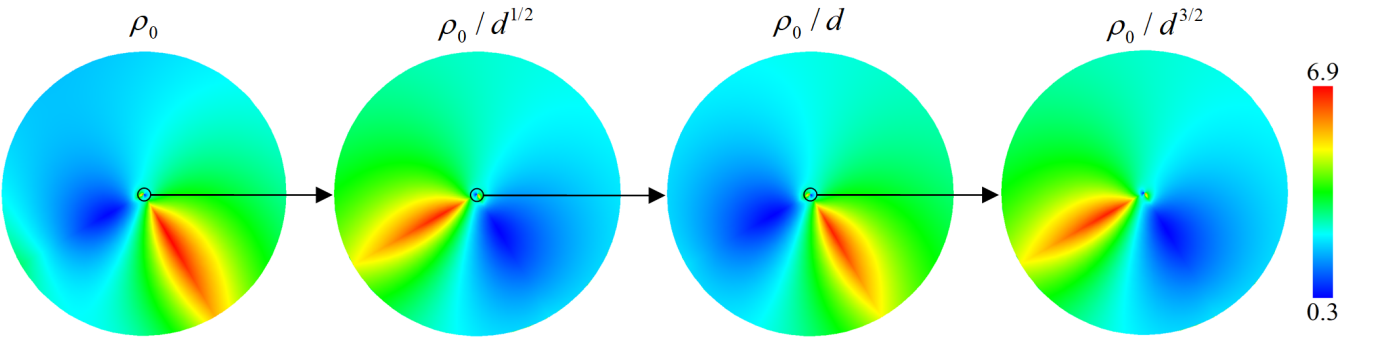

Fig. 13. The squared magnitude of the magnetic field $|H(x,y)|^2$ at the apex of the triangular diffraction grating No. 3 of Table 2 with $\varepsilon_s = -1.5 + 0i$ ($\text{Re}(\tau_1) = +0$), for several regions embedded in each other; the radius of the first region is $k_0\rho_0 = 0.3$, the divider $d^{1/2} = 23.3$. It can be seen that if the first and the third images are still different, then the second and the fourth are almost identical.

The radius of the region near the edge $k_0\rho_0 = 0.3$ in the first image of Fig. 13 is the same as in the Fig. 12. In each subsequent image of Fig. 13 the radius decreases by about $d^{1/2}=23.3$ times, the scale on the right is the same for all images. Despite of the reduction in scale, the field continues to change without decreasing the amplitude, without reaching a constant value. And if we wanted to represent such a distribution in the form of a complex Fourier series as in the C-method, or by using an orthogonal polynomials expansion as in the spectral element method, we would need more and more functions to take into account field changes.

To approximately represent the magnetic field near the top of the profile for small values of $\rho$, we restrict ourselves to the first terms of series (15) containing only $\tau_0$ and $\tau_1$:

$$H_z(\rho,\varphi) \approx \sum_{j,k=0,1} h_j \Phi_k(\varphi) \cdot (\rho/\rho_0)^{\tau_j} \delta_{k,j} = h_0 + h_1 \Phi_1(\varphi) \cdot (\rho/\rho_0)^{\tau_1}, \tag{33}$$

where we consider that $\tau_0 = 0$ and $\Phi_0(\varphi) = 1$.

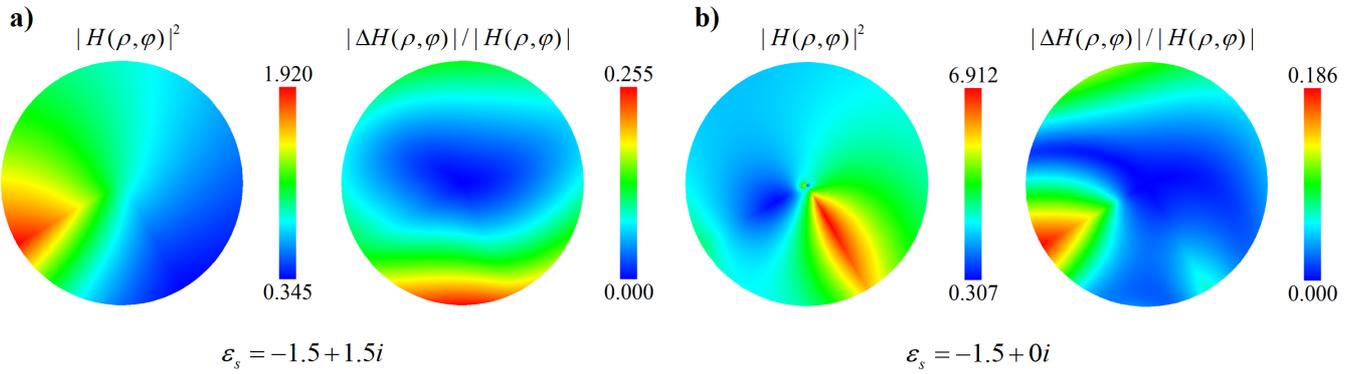

Fig. 14. The squared magnitude of the magnetic field $|H(x,y)|^2$ near the apex of the triangular profile inside the region $k_0\rho_0 = 0.3$ for two values of the dielectric constant of the grating material No. 3 of Table 2: a) $\varepsilon_s = -1.5+1.5i$, b) $\varepsilon_s = -1.5+0i$. The relative error of the expansion (33) that considers only the first terms with degrees $\tau_0$ and $\tau_1$ is shown beside.

Fig. 14 demonstrates the graphs of the relative error of such approximation Eq. (33) inside the region $k_0\rho_0 = 0.3$ near the edge at the apex of the triangular profile.

The value of $\tau_1$ in the general case is the imaginary quantity $\tau_1 = \alpha_1 + i\beta_1$, and the squared magnetic field can be written:

$$|H(\rho,\varphi)|^2 = |H_z(\rho,\varphi)|^2 \approx |h_0|^2 + (\rho/\rho_0)^{2\alpha_1}|h_1\Phi_1(\varphi)|^2 + 2(\rho/\rho_0)^{\alpha_1}\mathrm{Re}[h_0^* h_1 \Phi_1(\varphi) e^{i\beta_1 \ln(\rho/\rho_0)}]. \tag{34}$$

For a finite value of $\alpha_1 > 0$, the squared magnitude of the magnetic field for small $\rho$ converges to a constant $|h_0|^2$. If $\alpha_1 = +0$, $\beta_1 \neq 0$, then for small but finite values of $\rho \neq 0$ we have:

$$|H(\rho,\varphi)|^2 \approx |h_0|^2 + |h_1\Phi_1(\varphi)|^2 + 2\mathrm{Re}[h_0^* h_1 \Phi_1(\varphi) e^{i\beta_1 \ln(\rho/\rho_0)}] = a(\varphi) + b(\varphi)\cos(\beta_1 \ln\rho + \delta), \tag{35}$$

where $a(\varphi)$ and $b(\varphi)$ do not depend on $\rho$. The squared magnitude of the magnetic field in this case will oscillate with an increasing frequency when $\rho$ goes to zero. And for the radii $\rho_1/\rho_2 = \exp(-2\pi/\beta_1) \equiv d$ the picture will begin to repeat, as we saw in Fig. 13. The components of the electric field $E_\rho \sim \rho^{-1}\partial_\varphi H_z \sim \rho^{\tau_1-1}$, $E_\varphi \sim \partial_\rho H_z \sim \rho^{\tau_1-1}$ will oscillate with increasing amplitude.

Fig. 15 shows the squared magnitude of the magnetic field depending on the distance to the apex of the triangular grating (see Fig. 12) when moving along the profile from the left side to the top. The distance is plotted on a logarithmic scale. The dashed line shows the asymptotic approximation calculated by Eq. (34). In the case of $\varepsilon_s = -1.5+1.5i$, see Fig. 15a), when approaching the peak ($\rho \to 0$) the oscillations decay quickly, and the magnetic field converges to a constant.

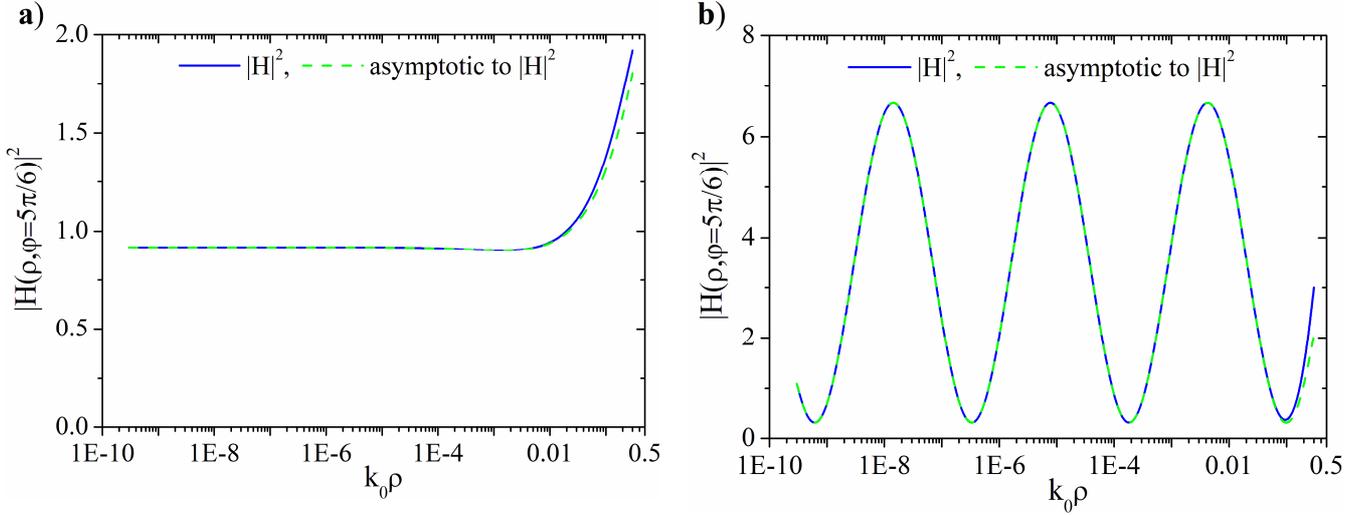

Fig. 15. The squared magnitude of the magnetic field depending on the distance to the apex of the triangular grating No. 3 of Table 2 (see Fig. 12), when moving along the profile line from the left side to the top for two values of the dielectric constant of the grating material: a) $\varepsilon_s = -1.5+1.5i$, b) $\varepsilon_s = -1.5+0i$. The dashed line shows the asymptotic approximation calculated by the Eq. (34).

In this case the field can be represented with any predetermined accuracy by using a finite number of basis functions, and the usual calculation methods work well here. However, the smaller the value of $\alpha_1 \equiv \text{Re}(\tau_1)$ in Eq. (34), the slower the oscillations damping will be, and the more functions will be needed to represent the solution at a given accuracy. This behavior is seen in Fig. 11b). For $\varepsilon_s = -1.5+0i$ the oscillations, in the case when the quantity $\rho$ decreases, persist with a constant amplitude and with an increasing frequency, therefore, it was not possible to obtain the result by conventional methods [7,8]. Nevertheless, the explicit calculation of the field near the edges, as shown in the present paper, copes with this problem.

## Conclusion

In this work we studied an approach which improves the accuracy and accelerates the convergence of numerical methods for solving the Maxwell's equations by means of taking into consideration the singularities of the electromagnetic field that arise near the geometric edges of scattering objects. Several algorithms to incorporate the singularities processing into methods for solving the Maxwell's equations in two-dimensional structures were analyzed. As implementation examples we used the analytical modal method and the spectral element method. In the first case, the resulting system of equations for the coefficients of the eigenmodes underwent changes. In the second case, for each region containing a singularity a Dirichlet-to-Neumann map operator was constructed and subsequently these regions were processed, like all the others.

In test examples, for which we used the calculation of the lamellar and the triangular diffraction gratings characteristics, a significant increase in accuracy and an acceleration of convergence was achieved. In the examples considered, the possibility to increase the type of convergence of spectral methods from algebraic to exponential or close to exponential was demonstrated. Diffraction efficiencies of the gratings, for which the conventional methods fail to converge due to the special values of permittivities, were calculated. In this work we confined ourselves to the two-dimensional case only; however, the method presented in the article can be extended to the three-dimensional case as well.

In conclusion we note, that the approach presented in the fourth paragraph of the paper, which employs the Dirichlet-to-Neumann map for joining subdomains, is universal and allows to easily em-bed the processing of singularities in various 2D multi-domain methods, regardless of what the numerical algorithm for solving Maxwell's equations is used inside individual domains. In addition to the spectral methods analyzed in the article, it can be the boundary integral method, the finite-difference method, the finite-element method, and etc.

## Appendix 1. Solution of the eigenvalue problem for angular functions $\Phi_k(\varphi)$ in some simple cases

The examples considered here have already been investigated previously in [9,10,24]. For completeness, we present the main results for simple cases, writing out the whole series of roots of the eigenvalues $\tau_k$, $\mathrm{Re}(\tau_k) \geq 0$, taking into consideration the degree of degeneracy, which is important for the correct writing of system (12). Note that when we write the final formulae for $\tau_k$ in this appendix, we do not sort $\tau_k$ by the value of the real part.

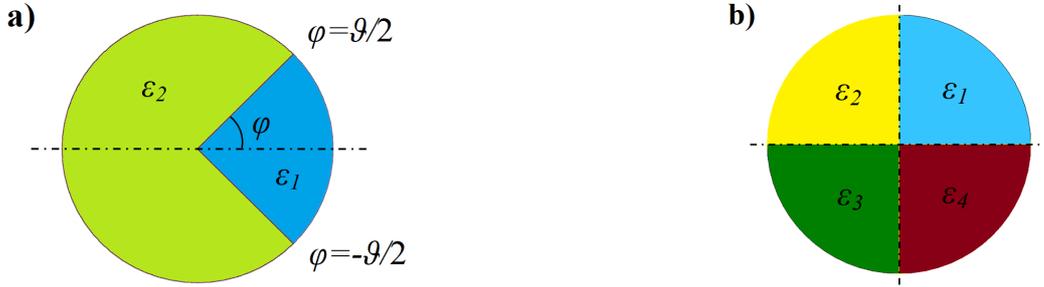

Fig. 16. Simple cases of wedge assemblies.

Let us at first solve the problem Eq. (6) in the case of a homogeneous infinite wedge with an angle $\vartheta$ made of a material with a permittivity $\varepsilon_1$ in a medium with a permittivity $\varepsilon_2$, Fig. 16a). We position the $x$ axis on the axis of symmetry, and we shall seek for the function $\Phi_k(\varphi)$ in the upper half-plane $0 \leq \varphi \leq \pi$, in the lower half it will be either antisymmetric or symmetric. Accordingly, at points $\varphi = 0, \pi$ either the function itself or its derivative must vanish. In the first case, the solution can be written as:

$$\Phi_k(\varphi) = \begin{cases} a_{1,k} \sin \tau_k \varphi, & 0 \leq \varphi < \vartheta/2 \\ a_{2,k} \sin \tau_k (\varphi - \pi), & \vartheta/2 \leq \varphi \leq \pi \end{cases}, \quad (36)$$

and in the second:

$$\Phi_k(\varphi) = \begin{cases} a_{1,k} \cos \tau_k \varphi, & 0 \leq \varphi < \vartheta/2 \\ a_{2,k} \cos \tau_k (\varphi - \pi), & \vartheta/2 \leq \varphi \leq \pi \end{cases}. \quad (37)$$

Using the boundary conditions Eq. (21) for the antisymmetric case Eq. (36) we have:

$$a_{1,k} \sin \tau_k \vartheta/2 - a_{2,k} \sin \tau_k (\vartheta/2 - \pi) = 0$$

$$\frac{a_{1,k} \tau_k}{\varepsilon_1} \cos \tau_k \vartheta/2 - \frac{a_{2,k} \tau_k}{\varepsilon_2} \cos \tau_k (\vartheta/2 - \pi) = 0 \quad . \quad (38)$$

Dividing the second line by the multipliers $\tau_k$, since the constant solution in the case of $\tau_0 = 0$ we already know, and equating the determinant of Eq. (38) to zero, we obtain the equation for $\tau_k$:

$$\varepsilon_1 \sin \tau_k \vartheta/2 \cdot \cos \tau_k (\vartheta/2 - \pi) - \varepsilon_2 \sin \tau_k (\vartheta/2 - \pi) \cdot \cos \tau_k \vartheta/2 = 0. \quad (39)$$

Similarly for the symmetric case:

$$\varepsilon_1 \cos \tau_k \vartheta/2 \cdot \sin \tau_k (\vartheta/2 - \pi) - \varepsilon_2 \cos \tau_k (\vartheta/2 - \pi) \cdot \sin \tau_k \vartheta/2 = 0. \tag{40}$$

It is convenient to rewrite equations (39), (40) in the form:
$$\pm \delta \cdot \sin \tau_k (\pi - \vartheta) = \sin \tau_k \pi, \quad \delta = (\varepsilon_1 - \varepsilon_2)/(\varepsilon_1 + \varepsilon_2), \tag{41}$$

where the "+" sign in front of $\delta$ corresponds to the antisymmetric case and "-" sign to the symmetric one. Note that replacing $\varepsilon_1 \leftrightarrow \varepsilon_2$ is identical to changing the sign "$\pm$" before $\delta$ in Eq. (41) by "$\mp$", thus the set $\{\tau_j\}$ in a simple triangular grating in Fig. 1b) is the same for the upper and for the lower edges, only the symmetry of the functions $\Phi_j(\varphi)$ for the corresponding $\tau_j$ will be different.
When $\vartheta = \pi/2$, the Eq. (41) will be:
$$\pm \delta \sin(\tau_k \pi/2) = \sin \tau_k \pi. \tag{42}$$

It can be noted that if $\tau_k$ is a solution of Eq. (42), then $\tau_k + 2n$, where $n$ is an integer, is also a solution and, considering the zero solution $\tau_0 = 0$, it is enough to find $\tau_k$ in the interval $0 < \text{Re}(\tau_k) \leq 2$, all the others are obtained by adding $2n, n = 1, 2, 3...$. Substituting $y = \tau_k \pi/2$, we obtain the equation $\pm \delta \sin y = 2 \sin y \cos y$, whose solutions will be:

$$\begin{aligned} &a)\ \cos y = \pm \delta/2 \Rightarrow \tau_1 = 4n \pm 2\pi^{-1} \arccos(+\delta/2), \tau_2 = 4n \pm 2\pi^{-1} \arccos(-\delta/2) \\ &b)\ \sin y = 0 \Rightarrow \tau_{3,4} = 2n \end{aligned} \tag{43}$$

The solutions $\tau = 4n \pm 2\pi^{-1} \arccos(+\delta/2)$ correspond to the antisymmetric case and $\tau = 4n \pm 2\pi^{-1} \arccos(-\delta/2)$ to the symmetric one. And there are two solutions $\tau = 2n$ in Eq. (43) b): one for the symmetric and one for the antisymmetric case. Thus, the whole set of solutions $\tau_k$, $\text{Re}(\tau_k) \geq 0$ can be written as:
$$\tau = \{0, 2\pi^{-1} \arccos(\delta/2), 2\pi^{-1} \arccos(-\delta/2), 2\} + 2n, \quad n = 0, 1, 2..., \tag{44}$$

Such a notation explicitly takes into account that the eigenvalues corresponding to $\tau = 2, 4, 6,...$ are doubly degenerate. Note that the middle terms in the brackets of expression (44) are not sorted by the value of the real part, as well as in the other formulae in this appendix, and to determine the $\tau_1$ characterizing the singularity it is necessary to choose one of them whose real part is minimal. In numerical calculation only the first $N_w$ terms of the series $\{\tau_k\}$ sorted in ascending order of the real parts of $\tau_k$ are used. In general case of arbitrary $\vartheta$ when $\tau_k$ is calculated numerically, for example, by method represented in Ref. [31], the symmetry of the eigenvectors corresponding to nondegenerate $\tau_k$ can be readily identified by substituting $\tau_k$ into Eq. (39) and Eq. (40) and comparing the absolute values of the right-hand sides. The values of the corresponding $\tau_k$ coefficients $a_{1,k}, a_{2,k}$ can be obtained from linear equation (38) for the symmetric and from similar equation for the antisymmetric case.
When $\vartheta = \pi \cdot l/m$, where $l, m = 1, 2, 3...$, we have:
$$\pm \delta \cdot \sin[\tau_k \pi (m-l)/m] = \sin \tau_k \pi. \tag{45}$$

It can be seen that it is enough to find solutions of Eq. (45) in the interval $0 < \text{Re}(\tau_k) \leq m$, all the others are obtained by adding $m \cdot n, n = 1, 2, 3...$. Moreover, between solutions $\tau = 0$ and $\tau = m$ there will be $2(m-1)$ solutions, and solutions $\tau = m \cdot n$ are twice degenerate, since they satisfy both the symmetric and the antisymmetric cases. For example, if $\vartheta = \pi/3$ then by substituting $y = \tau_k \pi/3$, we obtain the equations:
$$\pm \delta \sin 2y = \sin 3y \Leftrightarrow \pm 2\delta \sin(y) \cos(y) = 4 \sin(y) \cos(y)^2 - \sin(y), \tag{46}$$

whose solutions can be written in the form of six values $\tau_0,...,\tau_5$ with a shift of $3n$:

$$\tau = \{0, 3\pi^{-1}\arccos[(\pm\delta - \sqrt{\delta^2+4})/4], 3\pi^{-1}\arccos[(\pm\delta + \sqrt{\delta^2+4})/4], 3\} + 3n, \quad n = 0,1,2.... \quad (47)$$

And for $\vartheta = \pi \cdot 2/3$:

$$\tau = \{0, 3\pi^{-1}\arccos(-\sqrt{1\pm\delta}/2), 3\pi^{-1}\arccos(\sqrt{1\pm\delta}/2), 3\} + 3n, \quad n = 0,1,2.... \quad (48)$$

The next simple case is a combination of four right-angled wedges with permittivities $\varepsilon_l$, $l = 1..4$, Fig. 16b). The solution of Eq. (6) in each $l$-th section is:

$$\Phi_k(\varphi) = a_l e^{i\tau_k \varphi} + b_l e^{-i\tau_k \varphi}. \quad (49)$$

Using the conditions for the function and for the derivative at the four boundaries $\varphi_l = l\pi/2$, $l = 1..4$, we obtain a homogeneous system of 8 equations:

$$a_1 e^{i\tau\pi/2} + b_1 e^{-i\tau\pi/2} = a_2 e^{i\tau\pi/2} + b_2 e^{-i\tau\pi/2}, (a_1 e^{i\tau\pi/2} - b_1 e^{-i\tau\pi/2})/\varepsilon_1 = (a_2 e^{i\tau\pi/2} - b_2 e^{-i\tau\pi/2})/\varepsilon_2$$
$$\ldots \quad (50)$$
$$a_4 e^{i2\tau\pi} + b_4 e^{-i2\tau\pi} = a_1 + b_1, (a_4 e^{i2\tau\pi} - b_4 e^{-i2\tau\pi})/\varepsilon_4 = (a_1 - b_1)/\varepsilon_1$$

where we have divided the equations for derivatives by the factor $i \cdot \tau$, since we consider the constant solution at $\tau = 0$ already known. In order that the linear system (50) has a nonzero solution, it is necessary that its determinant must be equal to zero, from this condition the values of $\tau_k$ can be found. From Eq. (50) it follows that if $\tau_k$ satisfies the condition that the determinant of this equation is equal to zero, then the number $\tau_k + 2n$, where $n$ is an integer, also satisfy this condition. Thus, it is enough to find solutions in the interval $0 < \text{Re}(\tau_k) \leq 2$, all the others can be obtained by adding $2n, n = 1,2,3...$. Besides, if $\tau_k$ is a solution, then $2 - \tau_k$ is also solution, therefore, if there are solutions other than $0$ and $2n$, then such $\tau_k$ exists that $0 < \text{Re}(\tau_k) \leq 1$, and if $\tau_1 \neq 1$ then the singularity of the electric field components exists, which is proportional to $\rho^{\tau_1-1}$. By substituting $x = e^{i\tau_k\pi}$ and by equating determinant of Eq. (50) to zero we obtain:

$$e \cdot x^8 \left[ c \cdot (x^4+1) - 4d \cdot (x^3+x) - 2(c-4d) \cdot x^2 \right] = 0 \quad (51)$$
$$c = (\varepsilon_1+\varepsilon_2)(\varepsilon_2+\varepsilon_3)(\varepsilon_3+\varepsilon_4)(\varepsilon_4+\varepsilon_1), d = (\varepsilon_1\varepsilon_3 - \varepsilon_2\varepsilon_4)^2, e = (\varepsilon_1\varepsilon_2\varepsilon_3\varepsilon_4)^{-2}$$

Or, after simple conversions:

$$x^8(x-1)^2(x^2 - 2\alpha x + 1) = 0, \quad \alpha = (2d/c) - 1. \quad (52)$$

Zero solutions $x = 0$ have no physical meaning, two solutions with $x = 1$ correspond to $\tau_{3,4} = 2$, and solutions corresponding to the last bracket vanishing:

$$\alpha = (x^2+1)/2x = (x+x^{-1})/2 = \cos(\tau_k\pi), \quad (53)$$

are:

$$\tau_1 = \pi^{-1}\arccos(2d/c - 1) = 2\pi^{-1}\arccos(\sqrt{d/c}), \quad \tau_2 = 2 - 2\pi^{-1}\arccos(\sqrt{d/c}) \quad (54)$$

Thus, the whole set of numbers $\tau$ can be represented as:

$$\tau = \{0, 2\pi^{-1}\arccos(\sqrt{d/c}), 2 - 2\pi^{-1}\arccos(\sqrt{d/c}), 2\} + 2n, \quad n = 0,1,2.... \quad (55)$$

In the general case when all wedges in a wedge assembly have angles that are multiples of $\pi/m$, where $m$ is a positive integer, then it follows from the system of $4m$ equations for the continuity of functions $\Phi_k(\varphi)$ Eq. (49) and conditions on the derivative on $2m$ boundaries $\varphi_l = l\pi/m$, $l = 1..2m$:

$$a_1 e^{i\tau\pi/m} + b_1 e^{-i\tau\pi/m} = a_2 e^{i\tau\pi/m} + b_2 e^{-i\tau\pi/m}, (a_1 e^{i\tau\pi/m} - b_1 e^{-i\tau\pi/m})/\varepsilon_1 = (a_2 e^{i\tau\pi/m} - b_2 e^{-i\tau\pi/m})/\varepsilon_2$$
$$\ldots \tag{56}$$
$$a_{2m} e^{i2\tau\pi} + b_{2m} e^{-i2\tau\pi} = a_1 + b_1, (a_{2m} e^{i2\tau\pi} - b_{2m} e^{-i2\tau\pi})/\varepsilon_{2m} = (a_1 - b_1)/\varepsilon_1$$

that if $\tau_k$ satisfies the condition that the determinant of Eq. (56) is equal to zero, then the numbers $\tau_k + n \cdot m$, where $n$ is an integer, also satisfy this condition. Thus, it is enough to find solutions in the interval $0 < \text{Re}(\tau_k) \leq m$, all the rest are obtained by adding $n \cdot m$, $n = 1, 2, 3...$; besides, if $\tau_k$ is a solution then $m - \tau_k$ is also solution. It also follows that the number of distinct roots sorted in ascending order of the real part between the values $\tau_k + (n-1) \cdot m$ and $\tau_k + n \cdot m$ is the same for all $k$ and $n$. Assuming that the eigenvalues as quantities that have a physical meaning in modal methods continuously depend on the permittivity [1], it is easy to show that there will be exactly $2(m-1)$ solutions between the values $\tau = (n-1) \cdot m$ and $\tau = n \cdot m$. Indeed, in the case of uniform permittivity on the closed interval $[0, 2\pi]$ the unnormalized eigenfunctions of the problem:

$$\partial^2 \Phi_k(\varphi) = -\tau_k^2 \Phi_k(\varphi) \tag{57}$$

are $\{\cos(\tau_k \varphi), \sin(\tau_k \varphi)\}$ and all eigenvalues $\tau_k$, except for $\tau_0 = 0$, are doubly degenerate and represent a sequence of natural numbers: $\tau = 0, 1, 1, 2, 2, 3, 3, \ldots$. And there are $2(m-1)$ solutions between the values $\tau = (n-1) \cdot m$ and $\tau = n \cdot m$ in this case. By continuously changing the dielectric constant inside the wedges with angles that are multiples of $\pi/m$, one can see that the doubly degenerate eigenvalues $\tau_{2n \cdot m - 1} = \tau_{2n \cdot m} = n \cdot m$ remain unchanged, the first corresponds to the eigenfunction $\cos(\tau_{2n \cdot m - 1} \varphi)$ ($a_j = b_j = 1/2$ in Eq. (56)), the second corresponds to eigenfunction $\varepsilon(\varphi) \sin(\tau_{2n \cdot m} \varphi)$ ($a_j = -b_j = \varepsilon_j/(2\mathbf{i})$). Under the condition $\tau_{k'} = \tau_k + m$ between $\tau_{k'}$ and $\tau_k$ the number of roots, sorted by the real part, lying between the values $\tau = (n-1) \cdot m$ and $\tau = n \cdot m$, $n = 1, 2, \ldots$ must be the same. If the roots $\tau_k$ change continuously, that is they do not disappear and new ones do not arise, then the number of solutions with a real part between the values $(n-1) \cdot m$ and $n \cdot m$ will remain unchanged. Thus, we can write: $\tau_{k+2n \cdot m} = \tau_k + n \cdot m$.

In Appendix 2 we shall show that the fulfillment of the relation $\tau_{k'} = \tau_k + n \cdot m$ between the roots leads to the presence of the logarithms of $r$ in the series $R_{k,j}(r)$ Eq.(14), when the angles of the wedges are the product of rational numbers and $\pi$.

## Appendix 2. Calculation of radial functions R

### 2.1 Iterative calculation of radial functions R

We have to solve the system of differential equations (12) in the domain $0 \leq r \leq 1$, to obtain a set of functions $R_{k,j}(r)$ satisfying condition (10). In the numerical solution we restrict ourselves to the first functions $R_{k,j}(r)$, $k = 0..N_\Phi - 1$, $j = 0..N_w - 1$ corresponding to the roots $\tau_k$, sorted in ascending order of the real part. We shall solve the system of equations (12) iteratively, finding the new $l+1$ approximation $R_{k,j}^{[l+1]}$ by substituting of the previous $R_{k,j}^{[l]}$ into the right side of the system:

---

[1] The property of the continuous dependence of the roots $\tau_k$ on the dielectric constant is used in some numerical implementations of the analytical modal method to calculate the values of $\tau_k$, see, for example, [30].

$$\left[r\partial_r(r\partial_r) - \tau_k^2\right]R_{k,j}^{[l+1]}(r) = -r^2\gamma^2\sum_{s=0}^{N_\Phi-1}c_{k,s}R_{s,j}^{[l]}(r). \tag{58}$$

As a zeroth approximation we take the static solution:

$$R_{k,j}^{[0]}(r) = \delta_{k,j}R_j^0(r) = \delta_{k,j}r^{\tau_j}. \tag{59}$$

The approximations $R_{k,j}^{[l]}(r)$ can be calculated independently for each $j = 0..N_w - 1$, the difference is only in the initial approximation (59).

Solution of Eq. (58) is the sum of the general solution of the corresponding homogeneous system and the particular solution of the inhomogeneous one. As a solution of the homogeneous system that satisfies the finiteness condition at zero and Eq. (10), we take the static solution (59) at each iteration. At the first iteration the solution of equation (58) is

$$\left[r\partial_r(r\partial_r) - \tau_k^2\right]R_{k,j}^{[1]}(r) = -r^2\gamma^2 c_{k,j}r^{\tau_j} \Rightarrow R_{k,j}^{[1]}(r) = R_{k,j}^{[0]}(r) + \begin{cases} \gamma^2 c_{k,j}\dfrac{r^{\tau_j+2}}{\tau_k^2 - (\tau_j+2)^2}, & \tau_k \neq \tau_j + 2 \\ \gamma^2 c_{k,j}\dfrac{r^{\tau_j+2}(1 - 2\tau_k \ln r)}{4\tau_k^2}, & \tau_k = \tau_j + 2 \end{cases} \tag{60}$$

In Eq. (60) we obtained the sum of the zeroth approximation with a term containing the powers and logarithms of $r$ multiplied by $\gamma^2$. If we continue this process the result will be a series in even powers of $\gamma$. Besides,

$$R_{k,j}^{[l]}(r) = R_{k,j}^{[l-1]}(r) + \gamma^{2l}r^{\tau_j+2l}a_{k,j,l}(r), \tag{61}$$

where $a_{k,j,l}(r)$ consists of constants and possibly also logarithms of $r$. It can be seen from Eq. (60) that the power of $r$ on the right-hand side after each iteration increases by 2 and if there are $\tau_k = \tau_j + 2l$, where $l$ is some positive integer, then starting from the $l$-th approximation the term $R_{k,j}^{[l]}(r)$ will contain logarithms of $r$ and in the next iteration the logarithms will be already in all $R_{i,j}^{[l+1]}(r)$, $i = 0,1,2...$. As follows from Appendix 1, if the wedge assembly consists of wedges with angles that are multiples of $\pi/m$, where $m$ is a positive integer, or in other words, the angles are products of rational numbers by $\pi$, then there are roots $\tau_k = \tau_j + n \cdot m$, $n = 1,2,...$. Thus, logarithms appear in the approximation $R_{k,j}^{[l]}(r)$ as soon as $2l = n \cdot m$ is satisfied. And the functions $a_{k,j,i}(r)$ in Eq. (14) and Eq. (15) will contain the logarithms for $i \geq l$. When the angles are multiples of $\pi/2$, the logarithms in $R_{k,j}^{[l]}(r)$ appear starting from the first approximation, when the angles are multiples of $\pi/3$ starting from the third, etc.

At the next iterations we have to solve equations containing the right-hand side of the form:

$$\left[r\partial_r(r\partial_r) - \tau_k^2\right]u(r) = -r^2\gamma^2 c \cdot r^\alpha(\ln r)^\beta, \tag{62}$$

where the degree of the logarithm $\beta$ is an integer, $\beta < \text{Re}(\alpha)$. A particular solution to the equation with such a right-hand side can be calculated using formal integral representation:

$$u_{particular}(r) = -\gamma^2 C \cdot \begin{cases} \displaystyle\int_0^r \dfrac{\int_0^{r'}(r'')^{1+\alpha}(\ln r'')^\beta dr''}{r'} dr', & \tau_k = \tau_0 = 0 \\[2ex] \dfrac{1}{2\tau_k}\left[\dfrac{r^{\tau_k}(\ln r)^{\beta+1}}{\beta+1} - r^{-\tau_k}\int_0^r (r')^{2\tau_k-1}(\ln r')^\beta dr'\right], & \tau_k = \alpha+2 \\[2ex] \dfrac{1}{2\tau_k}\left[r^{\tau_k}\int_\sigma^r (r')^{\alpha-\tau_k+1}(\ln r')^\beta dr' - r^{-\tau_k}\int_0^r (r')^{\alpha+\tau_k+1}(\ln r')^\beta dr'\right], & \tau_k \neq \alpha+2 \end{cases}, \tag{63}$$

where in the third line for $\tau_k \neq \alpha + 2$ the lower limit of the first integral must be assigned as $\sigma = 0$ if $\mathrm{Re}(\tau_k) < \mathrm{Re}(\alpha) + 2$ and as $\sigma = +\infty$ otherwise. Integrals with a logarithm are calculated using the recurrence formula:

$$\int r^a (\ln r)^b dr = \frac{r^{a+1}(\ln r)^b}{a+1} - \frac{b}{a+1} \int r^a (\ln r)^{b-1} dr = r^{a+1} \sum_{q=0}^{b} \frac{(-1)^q b!}{(a+1)^{q+1}(b-q)!} \ln(r)^{b-q} + const. \quad (64)$$

So, Eq. (63) can be rewritten in expanded form:

$$u_{particular}(r) = \gamma^2 C r^{\alpha+2} \cdot \begin{cases} \sum_{q=0}^{\beta} \frac{(-1)^{q+1}(q+1)\beta!}{(\alpha+2)^{q+2}(\beta-q)!} (\ln r)^{\beta-q}, \tau_k = \tau_0 = 0 \\ \sum_{q=0}^{\beta+1} \frac{(-1)^{q+1}\beta!}{2^{q+1}(\alpha+2)^{q+1}(\beta+1-q)!} (\ln r)^{\beta+1-q}, \tau_k = \alpha+2 \\ \sum_{q=0}^{\beta} \frac{[(\alpha+2+\tau_k)^{q+1} - (\alpha+2-\tau_k)^{q+1}]\beta!}{2\tau_k[\tau_k^2 - (\alpha+2)^2]^{q+1}(\beta-q)!} (\ln r)^{\beta-q}, otherwise \end{cases} \quad (65)$$

Thus, the solution will look like the series:

$$R_{k,j}(r) = R_j^0(r)\left[\delta_{k,j} + \gamma^2 a_{k,j,1}(r) \cdot r^2 + \gamma^4 a_{k,j,2}(r) \cdot r^4 + \ldots\right], \quad (66)$$

where $a_{k,l}(r)$ are either constants independent of $r$, or if $\tau_k = \tau_j + 2l$ holds, where $l$ is a positive integer, then $a_{k,j,i}(r)$, $i \geq l$ consist of the sum of the constants and the logarithms in an integer power which is less than the power of multiplier $r$ in Eq. (66), so that $R_{k,j}(r \to 0) \to \delta_{k,j} R_j^0(r)$. The second option is realized when the angles of the wedges are the product of $\pi$ and rational numbers. As practical calculations show, for better accuracy the number $N_\Phi$ should be taken much greater than $N_w$, and on finishing the iterative process $R_{k,j}(r)$ can be truncated to a square matrix if necessary. The required number of functions $N_\Phi$ can be determined by taking the value several times larger than $N_w$ as the initial approximation, and then double $N_\Phi$ until the changes in the final calculation results become less than the specified error.

## 2.2 Convergence of the series $R_{k,j}(r)$

In this appendix we shall show that the series (66) converges absolutely for any finite values of $\gamma$, when the domain under consideration consists of wedges with angles multiples of $\pi/m$, where $m$ is a positive integer. We also briefly discuss the case of angles that are multiples of an irrational factor of $\pi$.

If the iterative process (61) is briefly written in the form: $R_{k,j}^{[l]}(r) = R_{k,j}^{[l-1]}(r) + u_{k,j,l}(r)$, $l = 1, 2 \ldots$, then the series (66) will look like:

$$R_{k,j}(r) = \delta_{k,j} R_j^0(r) + \sum_{l=1}^{\infty} u_{k,j,l}(r) = \sum_{l=0}^{\infty} u_{k,j,l}(r). \quad (67)$$

To prove the convergence of this series, below we will construct a majorizing series and show its convergence.

To write down analytical formulas for the terms of series (67), let us trace their evolution. During the first $L_1 - 1$ iterations, until the equality $\tau_{k_1} = \tau_j + 2L_1$ for some $\tau_{k_1}$ is satisfied, there are no logarithms in $R_{k,j}^{[l]}(r)$ and

$$R_{k,j}^{[l]}(r) = R_{k,j}^{[l-1]}(r) + \gamma^{2l} r^{t_j(l)} a_{k,j,l}(r) = R_{k,j}^{[l-1]}(r) + \gamma^{2l} r^{t_j(l)} A_{k,j}^{[l,0]}, \quad (68)$$

where we have introduced the notation $t_j(l) = \tau_j + 2l$ and $A_{k,j}^{[l,0]}$ are constants independent of $r$:

$$A_{k,j}^{[1,0]} = \frac{c_{k,j}}{\tau_k^2 - t_j(1)^2}, \quad A_{k,j}^{[2,0]} = \frac{1}{\tau_k^2 - t_j(2)^2} \sum_{s=0}^{\infty} c_{k,s} A_{s,j}^{[1,0]}, \ldots, A_{k,j}^{[l,0]} = \frac{1}{\tau_k^2 - t_j(l)^2} \sum_{s=0}^{\infty} c_{k,s} A_{s,j}^{[l-1,0]}. \quad (69)$$

The first upper index in $A_{k,j}^{[l,0]}$ corresponds to the iteration number and the second to power of the logarithm by which this coefficient will be multiplied. Note that $|t_j(l)| \geq 2l$ since $\mathrm{Re}(\tau_j) \geq 0$, $j = 0..\infty$ and the equality holds only for $j = 0, (\tau_0 = 0)$. For $1 < l < L_1$ the inequality $\tau_k \neq t_j(l)$ holds for all $k = 0..\infty$ and in the case under consideration, when the angles are proportional to the rational factor of the number $\pi$ it implies that [1] $\tau_{k+2n \cdot m} = \tau_k + n \cdot m$ and, for example, the first series in Eq. (69) can be written as:

$$\left| \sum_{s=0}^{\infty} \frac{c_{k,s} c_{s,j}}{\tau_s^2 - t_j^2} \right| \leq C^2 \sum_{s=0}^{\infty} \frac{1}{\left| \tau_s^2 - t_j^2 \right|} = C^2 \sum_{q=0}^{2m-1} \left[ \sum_{n=0}^{\infty} \frac{1}{\left| (\tau_q + n \cdot m)^2 - t_j^2 \right|} \right], \quad (70)$$

where $C = Max_{k,j} |c_{k,j}|$ is some finite number, since the functions $\Phi_k(\varphi)$ in Eq. (13) are continuous on a closed interval $[0, 2\pi]$ and, therefore, are bounded by the Weierstrass extreme value theorem and consequently the integral of their product on this interval is also bounded. Because the series in square brackets in Eq. (70) converges, the entire series (70) also converges. Similarly for other series in Eq. (69).

Since $\tau_{k+2n \cdot m} = \tau_k + n \cdot m$, the function

$$\hat{d}_j(l) = Min_{\substack{k=0..\infty, \\ \tau_k \neq t_j(l)}} |\tau_k - t_j(l)| \quad (71)$$

is periodic with period $\Delta l = m/2$ for even $m$ and $\Delta l = m$ otherwise. Let us define:

$$d_j = Min_{l=1..\infty} \hat{d}_j(l) = Min_{l=1..\Delta l} \hat{d}_j(l) \quad (72)$$

The finite length of the period $\Delta l$ is a fundamental feature of the case of angles that are multiples of the rational factor of $\pi$. This allows us to determine finite $d_j$ Eq. (72) and prove the convergence of the series (66) simply enough. Also in this case we can choose a continuous function $D_j(l)$ that grows with increasing $l$ no faster than logarithmic function, so that:

$$\sum_{\substack{k=0, \\ \tau_k \neq t_j(l)}}^{\infty} \frac{1}{\left| \tau_k^2 - t_j(l)^2 \right|} \leq \frac{\hat{D}_j(l)}{\hat{d}_j(l) \cdot |t_j(l)|} \leq \frac{D_j \cdot \ln(l+1)}{d_j \cdot |t_j(l)|}, \quad (73)$$

where $D_j > 0$ is some constant. This statement is easily proved by dividing (73) into the inner and outer sums as in Eq. (70) and replacing the inner sum over $n$ by a majorizing integral with appropriately chosen limits.

Note that if the condition $\tau_k \neq t_j(l)$, $k = 0..\infty$ was satisfied for any $l$ and there was a fixed value of $d_j$, then the convergence of series (66) for any finite $\gamma$ would be proved elementary, since the sequence $A_{k,j}^{[l,0]}$ for $l \gg 1$ asymptotically behaves like $|A_{k,j}^{[l,0]}| \sim C^l \left[ \prod_{q=1}^{l} \hat{D}_j(q) \right] / (d_j^l 2^l l!)$, which follows from (69), (73). And series (67) for $l \gg 1$ will be dominated by a converging series:

---

[1] If the number of roots between $\tau_k$ and $\tau_k + n \cdot m$, $n = 1, 2, \ldots$ differed from $2(m-1)$, the further proof would remain the same; it is only important for us that this number is permanent and does not depend on $n$ or $k$.

$$\tilde{u}_l = \gamma^{2l} C^l \left[ \prod_{q=1}^{l} \hat{D}_j(q) \right] / \left( d_j^l 2^l l! \right) . \tag{74}$$

As is known, series of the form $\sum_{l=0}^{\infty} \tilde{u}_l$ converge absolutely if starting from some index $l$, the ratio $|\tilde{u}_{l+1}/\tilde{u}_l|$ is less than some constant $q < 1$ (d'Alembert ratio test), since the series is dominated by a converging geometric progression with the common ratio $q$. For series (74), this criterion is satisfied for indexes $l > \gamma^2 C \cdot D_j \ln(l+2)/(2d_j)$. In our case, due to the fact that there are roots $\tau_k = t_j(L_\beta)$, $\beta = 1, 2 \ldots$, logarithms will appear in series (66), however, as we will see below, the asymptotic form of the terms of the series will remain approximately the same, and series (66) will converge for an arbitrary fixed $\gamma$.

When the equality $\tau_{k_1} = t_j(L_1)$, i.e. $\tau_{k_1} = \tau_j + 2L_1$ is valid for the first time for some $k_1$, then we get in the solution for the $k_1$-th row of system (58) (or in more rows, if there is a degeneracy) the logarithm. This will happen at least for $k_1 = j + 2n \cdot m$, so $\tau_j + 2L_1 = \tau_{k_1} = \tau_j + n \cdot m$, with $n = 1$ ($L_1 = m/2$) for even $m$ and $n = 2$ ($L_1 = m$) for odd $m$. The solution for $R_{k_1,j}^{[L_1]}(r)$ will be

$$R_{k_1,j}^{[L_1]}(r) = R_{k_1,j}^{[L_1-1]}(r) + \gamma^{2L_1} r^{t_j(L_1)} \left[ A_{k_1,j}^{[L_1,1]} d_j \ln r + A_{k_1,j}^{[L_1,0]} \right], \ \tau_{k_1} = t_j(L_1),$$

$$A_{k_1,j}^{[L_1,1]} = -\frac{1}{d_j \cdot 2 t_j(L_1)} \sum_{s=0}^{\infty} c_{k_1,s} A_{s,j}^{[L_1-1,0]}, \quad A_{k_1,j}^{[L_1,0]} = \frac{1}{2^2 t_j(L_1)^2} \sum_{s=0}^{\infty} c_{k_1,s} A_{s,j}^{[L_1-1,0]} \tag{75}$$

for all other equations of the infinite system (58), the solution remains in the form (68) with the coefficients:

$$A_{k,j}^{[L_1,0]} = \frac{1}{\tau_k^2 - t_j(L_1)^2} \sum_{s=0}^{\infty} c_{k,s} A_{s,j}^{[L_1-1,0]} = \frac{1}{[\tau_k - t_j(L_1)][\tau_k + t_j(L_1)]} \sum_{s=0}^{\infty} c_{k,s} A_{s,j}^{[L_1-1,0]}, \ k \neq k_1 \tag{76}$$

We specially added a factor $d_j$ to the denominator of the expression for the coefficient $A_{k_1,j}^{[L_1,1]}$ at the logarithm in Eq. (75) so that on the next iteration the coefficients $A_{k,j}^{[L_1+1,0]}$ and $A_{k,j}^{[L_1+1,1]}$ would be of the same order of magnitude in terms of $d_j$. Indeed, at the $L_1 + 1$-th iteration, equation (58) will have the form:

$$\left[ r \partial_r (r \partial_r) - \tau_k^2 \right] R_{k,j}^{[L_1+1]}(r) = -r^2 \gamma^2 \sum_{s=0}^{\infty} c_{k,s} R_{s,j}^{[L_1]}(r) =$$
$$= -r^2 \gamma^2 \left[ \sum_{s=0}^{\infty} c_{k,s} R_{s,j}^{[L_1-1]}(r) \right] - \gamma^{2(L_1+1)} r^{t_j(L_1+1)} \left[ c_{k,k_1} A_{k_1,j}^{[L_1,1]} d_j \ln r + \sum_{s=0}^{\infty} c_{k,s} A_{s,j}^{[L_1,0]} \right]. \tag{77}$$

The solution to Eq. (77) for $\tau_k \neq t_j(L_1 + 1)$ is

$$R_{k,j}^{[L_1+1]}(r) = R_{k,j}^{[L_1]}(r) + \gamma^{2(L_1+1)} r^{t_j(L_1+1)} \left[ A_{k,j}^{[L_1+1,1]} d_j \ln r + A_{k,j}^{[L_1+1,0]} \right], \tag{78}$$

where

$$A_{k,j}^{[L_1+1,0]} = \begin{cases} -\dfrac{1}{t_j(L_1+1)^2} \left[ \sum_{s=0}^{\infty} c_{0,s} A_{s,j}^{[L_1,0]} - c_{0,k_1} A_{k_1,j}^{[L_1,1]} \dfrac{2 d_j}{t_j(L_1+1)} \right], & k = 0 \\[2mm] \dfrac{1}{\tau_k^2 - t_j(L_1+1)^2} \left[ \sum_{s=0}^{\infty} c_{k,s} A_{s,j}^{[L_1,0]} + c_{k,k_1} A_{k_1,j}^{[L_1,1]} \dfrac{d_j t_j(L_1+1)}{\tau_k^2 - t_j(L_1+1)^2} \right], & k > 0 \end{cases} \tag{79}$$

$$A_{k,j}^{[L_1+1,1]} = \frac{1}{\tau_k^2 - t_j(L_1+1)^2} c_{k,k_1} A_{k_1,j}^{[L_1,1]}$$

The sum in the expression (79) for the coefficient $A_{k,j}^{[L_1+1,0]}$ is

$$\sum_{s=0}^{\infty} c_{k,s} A_{s,j}^{[L_1,0]} = \sum_{\substack{s=0, \\ s \neq k_1}}^{\infty} \frac{c_{k,s}}{[\tau_s - t_j(L_1)][\tau_s + t_j(L_1)]} U_{s,j} + \frac{c_{k,k_1}}{2^2 t_j(L_1)^2} U_{k_1,j} , \qquad (80)$$

and the product in the case of coefficient $A_{k,j}^{[L_1+1,1]}$ is

$$c_{k,k_1} A_{k_1,j}^{[L_1,1]} = -\frac{1}{d_j \cdot 2 t_j(L_1)} c_{k,k_1} U_{k_1,j} , \qquad (81)$$

where for brevity we introduced the notation $U_{s,j} = \sum_{q=0}^{\infty} c_{s,q} A_{q,j}^{[L_1-1,0]}$. Taking into account definitions (71), (72), it can be seen that expressions (80) and (81) are of the same order of magnitude in terms of $d_j$.

If at this iteration the equality $\tau_{k_2} = t_j(L_1+1)$ is satisfied for some index $k_2$, the logarithm to the second power will appear in the row $k_2$. Otherwise, until the equality $\tau_{k_2} = t_j(L_2)$ for some $k_2$ is satisfied at the next iterations, the logarithm of the first degree will be present on the right side of each line of Eq. (58), and for iteration $L_1 < l < L_2$ the solution will be written as:

$$R_{k,j}^{[l]}(r) = R_{k,j}^{[l-1]}(r) + \gamma^{2(l)} r^{t_j(l)} \left[ A_{k,j}^{[l,1]} d_j \ln r + A_{k,j}^{[l,0]} \right] , \qquad (82)$$

where

$$A_{k,j}^{[l,0]} = \begin{cases} -\dfrac{1}{t_j(l)^2} \sum_{s=0}^{\infty} c_{k,s} \left[ A_{s,j}^{[l-1,0]} - A_{s,j}^{[l-1,1]} d_j \dfrac{2}{t_j(l)} \right], & k = 0 \\[2mm] \dfrac{1}{\tau_k^2 - t_j(l)^2} \sum_{s=0}^{\infty} c_{k,s} \left[ A_{s,j}^{[l-1,0]} + A_{s,j}^{[l-1,1]} d_j \dfrac{t_j(l)}{\tau_k^2 - t_j(l)^2} \right], & k > 0 \end{cases} \qquad (83)$$

$$A_{k,j}^{[l,1]} = \frac{1}{\tau_k^2 - t_j(l)^2} \sum_{s=0}^{\infty} c_{k,s} A_{s,j}^{[l-1,1]}$$

When the equality $\tau_{k_2} = t_j(L_2)$ is realized for some $k_2$, we get in the solution for the $k_2$-th row of system (58) (or in more rows, if there is a degeneracy) the squared logarithm:

$$R_{k_2,j}^{[L_2]}(r) = R_{k_2,j}^{[L_2-1]}(r) + \gamma^{2L_2} r^{t_j(L_2)} \sum_{s=0}^{\infty} c_{k_2,s} \left\{ A_{s,j}^{[L_2-1,0]} \left[ -\frac{\ln r}{2 t_j(L_2) \cdot 1} + \frac{1}{2^2 t_j(L_2)^2} \right] + \right.$$

$$\left. + A_{s,j}^{[L_2-1,1]} d_j \left[ -\frac{(\ln r)^2}{2 t_j(L_2) \cdot 2} + \frac{\ln r}{2^2 t_j(L_2)^2} - \frac{1}{2^3 t_j(L_2)^3} \right] \right\}, \tau_{k_2} = t_j(L_2) \qquad (84)$$

or

$$R_{k_2,j}^{[L_2]}(r) = R_{k_2,j}^{[L_2-1]}(r) + \gamma^{2L_2} r^{t_j(L_2)} \left[ A_{k_2,j}^{[L_2,2]} d_j^2 \frac{(\ln r)^2}{2 \cdot 1} + A_{k_2,j}^{[L_2,1]} d_j \frac{\ln r}{1} + A_{k_2,j}^{[L_2,0]} \right] , \qquad (85)$$

where

$$A_{k_2,j}^{[L_2,2]} = -\frac{1}{d_j 2 t_j(L_2)} \sum_{s=0}^{\infty} c_{k_2,s} A_{s,j}^{[L_2-1,1]}, \quad A_{k_2,j}^{[L_2,1]} = \frac{1}{d_j 2 t_j(L_2)} \sum_{s=0}^{\infty} c_{k_2,s} \left[ -A_{s,j}^{[L_2-1,0]} + \frac{A_{s,j}^{[L_2-1,1]} d_j}{2 t_j(L_2)} \right],$$

$$A_{k_2,j}^{[L_2,0]} = \frac{1}{2^2 t_j(L_2)^2} \sum_{s=0}^{\infty} c_{k_2,s} \left[ A_{s,j}^{[L_2-1,0]} - \frac{A_{s,j}^{[L_2-1,1]} d_j}{2 t_j(L_2)} \right] \qquad (86)$$

For all other lines of the solution everything will be as before, in accordance with Eq. (82). Here we also added $d_j$ to the denominator of the coefficient $A_{k_2,j}^{[L_2,2]}$ in Eq. (86), so that $A_{k,j}^{[L_2+1,0]}$, $A_{k,j}^{[L_2+1,1]}$ and $A_{k,j}^{[L_2+1,2]}$ would be of the same order of magnitude in terms of $d_j$. In addition, in Eq. (85) we transfer the factor $1/2$ from the coefficient $A_{k_2,j}^{[L_2,2]}$ to the logarithm.

If we further continue this process, then in each iteration $L_\beta$, when $\tau_{k_\beta} = t_j(L_\beta)$ is satisfied for some $k_\beta$, in the solution for the $k_\beta$-th row of system (58) (or in more rows, if there is a degeneracy), the logarithm to power $\beta$ is added and the factor of this logarithm decreases with increasing $\beta$:

$$\frac{\gamma^{2L_\beta} r^{t_j(L_\beta)} d_j^{\beta-1}}{2t_j(L_\beta)\cdot(\beta-1)!} \sum_{s=0}^{\infty} c_{k_\beta,s} A_{s,j}^{[L_\beta-1,\beta-1]} \left[ -\frac{(\ln r)^\beta}{\beta} + r^{-2t_j(L_\beta)} \int_0^r (r')^{2t_j(L_\beta)-1} (\ln r')^{\beta-1} dr' \right]. \qquad (87)$$

The factorial in the denominator of Eq. (87) arises due to the fact that each logarithm, when it appears in Eq. (63), is divided by its power. And this factorial balances out the factorials arising in the process of integration in numerators, see Eq. (65). The complete expression for the solution $R_{k,j}^{[L_\beta]}(r)$ at iteration $L_\beta$ is

$$R_{k,j}^{[L_\beta]}(r) = R_{k,j}^{[L_\beta-1]}(r) + \gamma^{2L_\beta} r^{t_j(L_\beta)} \sum_{q=0}^{Q} \frac{d_j^q}{q!} A_{k,j}^{[L_\beta,q]} (\ln r)^q, \quad Q = \begin{cases} \beta, & \tau_k = t_j(L_\beta) \\ \beta-1, & otherwise \end{cases}, \qquad (88)$$

where the coefficients $A_{k,j}^{[L_\beta,q]}$ are defined recurrently as nested series:

$$A_{k,j}^{[L_\beta,q]} = \begin{cases} \sum_{s=0}^{\infty} c_{k,s} \sum_{p=0}^{\beta-1-q} d_j^p \dfrac{(-1)^{p+1}(p+1)}{t_j(L_\beta)^{p+2}} A_{s,j}^{[L_\beta-1,p+q]}, & \tau_k = \tau_0 = 0 \\[2mm] \sum_{s=0}^{\infty} c_{k,s} \sum_{p=Max(1-q,0)}^{\beta-q} d_j^{p-1} \dfrac{(-1)^{p+1}}{2^{p+1} t_j(L_\beta)^{p+1}} A_{s,j}^{[L_\beta-1,p+q-1]}, & \tau_k = t_j(L_\beta) \\[2mm] \sum_{s=0}^{\infty} c_{k,s} \sum_{p=0}^{\beta-1-q} d_j^p \dfrac{[t_j(L_\beta)+\tau_k]^{p+1} - [t_j(L_\beta)-\tau_k]^{p+1}}{2\tau_k [\tau_k^2 - t_j(L_\beta)^2]^{p+1}} A_{s,j}^{[L_\beta-1,p+q]}, & otherwise \end{cases} \qquad (89)$$

Note that for $|\tau_k| \ll |t_j(L_\beta)|$, the third row in Eq. (89) turns into an expression for the case $\tau_{k=0} = 0$ in the first row of Eq. (89). Also, when calculating the coefficient $A_{s,j}^{[L_\beta,q]}$, only the coefficients $A_{s,j}^{[L_\beta-1,Q]}$ with $Q \geq q$ are involved in the first and third lines of Eq. (89), and with $Q \geq Max(0,q-1)$ in the second line, when $\tau_k = t_j(L_\beta)$.

The series in formula (88) is dominated by the Maclaurin series for the exponent:

$$|u_{k,j,L_\beta}(r)| = \gamma^{2L_\beta} \left| r^{t_j(L_\beta)} \sum_{q=0}^{Q} \frac{d_j^q}{q!} A_{k,j}^{[L_\beta,q]} (\ln r)^q \right| \leq \gamma^{2L_\beta} \left| r^{t_j(L_\beta)} \right| Max_q \left| A_{k,j}^{[L_\beta,q]} \right| \sum_{q=0}^{\infty} \frac{d_j^q}{q!} |\ln r|^q \leq$$

$$\leq \gamma^{2L_\beta} \left| r^{t_j(L_\beta)} \right| Max_q \left| A_{k,j}^{[L_\beta,q]} \right| \exp[-d_j \ln r] = \gamma^{2L_\beta} r^{Re[t_j(L_\beta)]-d_j} Max_q \left| A_{k,j}^{[L_\beta,q]} \right| \leq \gamma^{2L_\beta} Max_q \left| A_{k,j}^{[L_\beta,q]} \right| \qquad (90)$$

where we took into account that $|\ln r| = -\ln r$, since $0 \leq r \leq 1$. By definition $Re[t_j(L_\beta)] \geq d_j$, so there is no singularity in Eq. (90). Thus, the convergence of series (88) is completely determined by the behavior of the coefficients $A_{k,j}^{[l,q]}$. If with an increase in the iteration number $l$ they change according to the power law: $|A_{k,j}^{[l,q]}| \sim \alpha^l$, where $\alpha$ is a constant, then the parameter $\gamma = 2\pi(\rho_0/\lambda)$ characterizing the radius of convergence $\rho_0$ is bounded above by some value $\gamma_{max} \sim \alpha^{-1/2}$. If with an increase in $l$ the coefficients $A_{k,j}^{[l,q]}$ decrease faster than according to a power law, for example, inversely proportional to the factorial $l$, then the series will converge for any finite $\gamma$. Before embarking on an analytical study of series (89), let us carry out a numerical experiment. If the first case $|A_{k,j}^{[l,q]}| \sim \alpha^l$ is realized, then the value of $\ln|A_{k,j}^{[l+1,q]}/A_{k,j}^{[l,q]}|$ with an increase in $l$ will come to a constant $\ln\alpha$. In the second case, this value will decrease indefinitely; so, for example, for $|A_{k,j}^{[l,q]}| \sim \alpha^l/l!$, it will be proportional to $\ln[\alpha/(l+1)]$. In Fig. 17, the solid blue line shows the dependence of $\ln\left(Max_{k,q}|A_{k,j}^{[l+1,q]}|/Max_{k,q}|A_{k,j}^{[l,q]}|\right)$ versus the iteration number $l$ at $j=0$ for the triangular diffraction grating No. 3 of Table 2, next to it the function $\ln[\alpha/(l+1)]$ is depicted by the green dashed line. A similar decreasing dependence can be obtained for other examples and numbers $j$, indicating that the second option is actually realized.

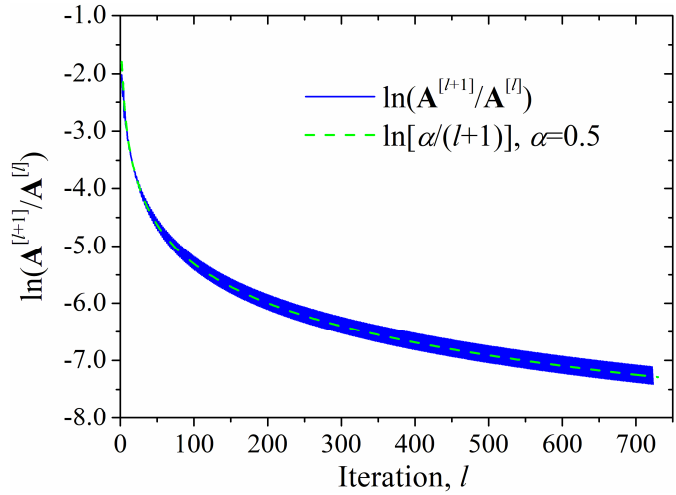

Fig. 17. The solid blue line shows the dependence of $\ln(\mathbf{A}^{[l+1]}/\mathbf{A}^{[l]})$, where $\mathbf{A}^{[l]} = Max_{k,q}|A_{k,j}^{[l,q]}|$ versus the iteration number $l$ at $j=0$ for the triangular diffraction grating No. 3 of Table 2, next to it the function $\ln[\alpha/(l+1)]$, $\alpha=0.5$ is depicted by the green dashed line.

Let us at first estimate the sum in the third line Eq. (89):

$$|A_{k,j}^{[L_\beta,q]}| \leq C \cdot F_j(L_\beta) \sum_{p=0}^{\beta-1-q} d_j^p \left|\frac{[t_j(L_\beta)+\tau_k]^{p+1} - [t_j(L_\beta)-\tau_k]^{p+1}}{2\tau_k[\tau_k^2 - t_j(L_\beta)^2]^{p+1}}\right| = C \cdot F_j(L_\beta) \sum_{p=0}^{\beta-1-q} d_j^p |T_{k,j}^{[L_\beta,p]}|, \quad (91)$$

where we denote $F_j(L_\beta) \equiv Max_Q \sum_{s=0}^{\infty} |A_{s,j}^{[L_\beta-1,Q]}|$. The absolute value of $T_{k,j}^{[L_\beta,p]}$ can be estimated as:

$$|T_{k,j}^{[L_\beta,p]}| = \frac{1}{2|\tau_k|}\left|\frac{1}{[\tau_k-t_j(L_\beta)]^{p+1}} - \frac{(-1)^{p+1}}{[\tau_k+t_j(L_\beta)]^{p+1}}\right| \leq \begin{cases} \dfrac{1}{|\tau_k||t_j(L_\beta)-\tau_k|^{p+1}}, & |\tau_k - t_j(L_\beta)| \leq |\tau_k + t_j(L_\beta)| \\ \dfrac{1}{|\tau_k||t_j(L_\beta)+\tau_k|^{p+1}}, & otherwise \end{cases} \quad (92)$$

In the case of $|\tau_k - t_j(L_\beta)| \leq |\tau_k + t_j(L_\beta)|$, taking into account that $d_j \leq |t_j(L_\beta)-\tau_k|$, for the sum in Eq. (91) we obtain:

$$\sum_{p=0}^{\beta-q-1} d_j^p |T_{k,j}^{[L_\beta,p]}| \leq \sum_{p=0}^{\beta-q-1} \frac{d_j^p}{|\tau_k||t_j(L_\beta)-\tau_k|^{p+1}} = \frac{1}{|\tau_k||t_j(L_\beta)-\tau_k|}\sum_{p=0}^{\beta-q-1}\frac{d_j^p}{|t_j(L_\beta)-\tau_k|^p} \leq \frac{\beta-q}{|\tau_k||t_j(L_\beta)-\tau_k|}. \quad (93)$$

And it can be written:
$$\left|A_{k,j}^{[L_\beta,q]}\right| \le C \cdot F_j(L_\beta) \frac{\beta-q}{|\tau_k||t_j(L_\beta)-\tau_k|}, \quad |\tau_k - t_j(L_\beta)| \le |\tau_k + t_j(L_\beta)|. \tag{94}$$

If $|\tau_k - t_j(L_\beta)| > |\tau_k + t_j(L_\beta)|$, then since $d_j \le |t_j(L_\beta)+\tau_k|$, we obtain a relation that is completely analogous to equation (93):

$$\sum_{p=0}^{\beta-q-1} d_j^p \left|T_{k,j}^{[L_\beta,p]}\right| \le \sum_{p=0}^{\beta-q-1} \frac{d_j^p}{|\tau_k||t_j(L_\beta)+\tau_k|^{p+1}} = \frac{1}{|\tau_k||t_j(L_\beta)+\tau_k|} \sum_{p=0}^{\beta-q-1} \frac{d_j^p}{|t_j(L_\beta)+\tau_k|^p} \le \frac{\beta-q}{|\tau_k||t_j(L_\beta)+\tau_k|}. \tag{95}$$

For the case $k=0$, $(\tau_0=0)$, taking into account that $d_j \le |t_j(1)-\tau_j|=2$ and $|t_j(L_\beta)| \ge 2\beta$, we have:

$$\sum_{p=0}^{\beta-1-q} d_j^p \frac{(p+1)}{|t_j(L_\beta)|^{p+2}} \le \frac{1}{|t_j(L_\beta)|d_j} \sum_{p=0}^{\beta-1-q} \frac{d_j^{p+1}(p+1)}{(2\beta)^{p+1}} \le \frac{1}{|t_j(L_\beta)|d_j} \sum_{p=0}^{\beta-1-q} \frac{(p+1)}{\beta^{p+1}} \le \frac{1}{|t_j(L_\beta)|d_j} \le \frac{\beta-q}{|t_j(L_\beta)|d_j} \tag{96}$$

For definiteness, let us analyze the case, when in each iteration $l$ the equality $\tau_{k_l} = t_j(l)$ is realized for some index $k_l$ and the maximum power of the logarithm in Eq. (88) increases by unity (this is the case when the angles of wedges are proportional to $\pi/2$). For other cases everything is done in the same way, we will discuss the difference below. For the second line of Eq. (89), when $\tau_{k_\beta} = t_j(L_\beta)$, the upper bound of the absolute value of $A_{k_\beta,j}^{[L_\beta,q]}$ will be

$$\left|A_{k_\beta,j}^{[L_\beta,q]}\right| \le \frac{C \cdot F_j(L_\beta)}{2|t_j(L_\beta)|d_j} \sum_{p=Max(1-q,0)}^{\beta-q} \frac{d_j^p}{2^p |t_j(L_\beta)|^p} \le \frac{C \cdot F_j(L_\beta)}{|t_j(L_\beta)|d_j}, \quad q \le \beta. \tag{97}$$

Passing to the next $(L_\beta+1)$-th iteration, let us estimate the sums over $s$ in Eq. (89), considering Eq. (94)-(96):

$$\left|\sum_{\substack{s=0 \\ s \ne k_\beta}}^{\infty} c_{k,s} A_{s,j}^{[L_\beta,q]}\right| \le C \cdot F_j(L_\beta) \cdot C \sum_{s=0}^{\infty} \left|A_{s,j}^{[L_\beta,q]}\right| = C^2 F_j(L_\beta) \tilde{D}_j(L_\beta) \frac{\beta-q}{|t_j(L_\beta)|d_j}, \quad q < \beta, \tag{98}$$

where the function $\tilde{D}_j(l) > 0$ increases no faster than logarithmic function. For the case $\tau_{k_\beta} = t_j(L_\beta)$, there will be one term (or more, if there is a degeneracy) for $s = k_\beta$ from Eq. (97):

$$\left|c_{k,k_\beta} A_{k_\beta,j}^{[L_\beta,q]}\right| \le C^2 F_j(L_\beta) \left[|t_j(L_\beta)|d_j\right]^{-1}, \quad q \le \beta. \tag{99}$$

Substituting Eq. (98) and Eq. (99) into the analogue of the expression in the third line of Eq. (89) but for the iteration $(L_\beta+1)$, under the condition $|\tau_k - t_j(L_\beta+1)| \le |\tau_k + t_j(L_\beta+1)|$ we obtain:

$$\left|A_{k,j}^{[L_\beta+1,q]}\right| \le \frac{C^2 F_j(L_\beta)}{|t_j(L_\beta)|d_j} \sum_{p=0}^{\beta-q} \frac{d_j^p \{\tilde{D}_j(L_\beta)[\beta-(p+q)]+\eta\}}{|\tau_k||t_j(L_\beta+1)-\tau_k|^{p+1}} \le$$
$$\le \frac{1}{|\tau_k||t_j(L_\beta+1)-\tau_k|} \frac{C^2 F_j(L_\beta)}{|t_j(L_\beta)|d_j} \left[\tilde{D}_j(L_\beta) \frac{(\beta-q)(\beta+1-q)}{2} + \eta \cdot (\beta+1-q)\right], \quad q \le \beta \tag{100}$$

where $\eta$ is the order of degeneracy for $\tau_k = t_j(L_\beta)$. For the case: $|\tau_k - t_j(L_\beta+1)| > |\tau_k + t_j(L_\beta+1)|$, the expression $|t_j(L_\beta+1)+\tau_k|$ would be contained in the denominator of Eq. (100) instead of $|t_j(L_\beta+1)-\tau_k|$. A similar expression can be obtained for the case $\tau_k = \tau_0 = 0$, with the only difference

that instead of the expression $|\tau_k||t_j(L_\beta+1)-\tau_k|$ the denominator will contain $|t_j(L_\beta+1)|d_j$. The same differences will be observed in the next iterations, so we will not focus further on the cases $|\tau_k - t_j(l)| > |\tau_k + t_j(l)|$ and $\tau_k = \tau_0 = 0$.

Note that the term with the factor $(\beta-q)(\beta+1-q)$ in Eq. (100) contributes to $A_{s,j}^{[L_\beta+1,q]}$ only for $q < \beta$, in accordance with the remark immediately after Eq. (89).

For the analogue of the expression in the second line of Eq. (89) corresponding to $\tau_{k_{\beta+1}} = t_j(L_\beta+1)$ we have:

$$\left|A_{k_{\beta+1},j}^{[L_\beta+1,q]}\right| \leq \frac{C^2 F_j(L_\beta)}{|t_j(L_\beta)|d_j}\left\{\sum_{p=\text{Max}(1-q,0)}^{\beta-q} \frac{d_j^{p-1}\tilde{D}_j(L_\beta)[\beta-(p+q)]}{2^{p+1}|t_j(L_\beta+1)|^{p+1}} + \right.$$

$$\left. + \sum_{p=\text{Max}(1-q,0)}^{\beta-q+1} \frac{d_j^{p-1}\eta}{2^{p+1}|t_j(L_\beta+1)|^{p+1}}\right\} \leq \frac{C^2 F_j(L_\beta)}{|t_j(L_\beta)t_j(L_\beta+1)|d_j^2}\left[\frac{\tilde{D}_j(L_\beta)}{2}\frac{(\beta-q)(\beta+1-q)}{2}+\eta\right] \quad (101)$$

In the $(L_\beta+2)$-th iteration the sums over $s$ in case of $q \leq \beta$ can be estimated as:

$$\left|\sum_{s=0}^\infty c_{k,s} A_{s,j}^{[L_\beta+1,q]}\right| \leq \frac{C^3 F_j(L_\beta)\tilde{D}_j(L_\beta+1)}{|t_j(L_\beta)t_j(L_\beta+1)|d_j^2}\left[\tilde{D}_j(L_\beta)\frac{(\beta-q)(\beta+1-q)}{2}+\eta(\beta+1-q)\right], \quad (102)$$

where we have included the term proportional to $\tilde{D}_j(L_\beta)$ from Eq. (101) for $A_{k_{\beta+1},j}^{[L_\beta+1,q]}$.

For the case $\tau_{k_{\beta+1}} = t_j(L_\beta+1)$ and $q = \beta+1$, there will be one term (or more, if there is a degeneracy) for $s = k_{\beta+1}$ from Eq. (101):

$$\left|c_{k,k_{\beta+1}} A_{k_{\beta+1},j}^{[L_\beta+1,\beta+1]}\right| \leq \frac{C^3 F_j(L_\beta)}{|t_j(L_\beta)t_j(L_\beta+1)|d_j^2}\eta. \quad (103)$$

Substituting Eq. (102) and Eq. (103) into counterpart of the third line of Eq. (89) for the iteration $(L_\beta+2)$, we obtain:

$$\left|A_{k,j}^{[L_\beta+2,q]}\right| \leq \frac{C^3 F_j(L_\beta)}{|t_j(L_\beta)t_j(L_\beta+1)|d_j^2} \times$$

$$\times \sum_{p=0}^{\beta+1-q} \frac{d_j^p\left(\tilde{D}_j(L_\beta+1)\{\tilde{D}_j(L_\beta)[\beta-(p+q)][\beta+1-(p+q)]+\eta[\beta+1-(p+q)]\}+\eta^2\right)}{|\tau_k||t_j(L_\beta+2)-\tau_k|^{p+1}} \leq$$

$$\leq \frac{1}{|\tau_k||t_j(L_\beta+2)-\tau_k|}\frac{C^3 F_j(L_\beta)}{|t_j(L_\beta)t_j(L_\beta+1)|d_j^2}\left[\tilde{D}_j(L_\beta+1)\tilde{D}_j(L_\beta)\frac{(\beta-q)(\beta+1-q)(\beta+2-q)}{3!}+\right.$$

$$\left.+\eta\hat{D}_j(L_\beta+1)\frac{(\beta+1-q)(\beta+2-q)}{2!}+\eta^2(\beta+2-q)\right] \quad (104)$$

When calculating the sum over index $p$, we used the formula:

$$\sum_{p=0}^{\beta+n-q}[\beta-(p+q)][\beta+1-(p+q)]\times\ldots\times[\beta+n-(p+q)] = \frac{[\beta-q][\beta+1-q]\times\ldots\times[\beta+n+1-q]}{n+2}, \quad (105)$$

which can be obtained from the relation for the sums of binomial coefficients: $\sum_{k=0}^m \binom{n+k}{n} = \binom{n+m+1}{n+1}$ [44]. And we assume that the order of degeneracy $\eta$ is the same for $\tau_k = t_j(L_\beta+1)$, if this is not the case, then we choose as $\eta$ the maximum order of degeneracy. Continuing this process further, after $L_\beta + N$ iterations for the case $\tau_{k_{\beta+N}} \neq t_j(L_\beta + N)$, $|\tau_k - t_j(L_\beta + N)| > |\tau_k + t_j(L_\beta + N)|$, $\tau_k \neq 0$ we get:

$$|A_{k,j}^{[L_\beta+N,q]}| \leq \frac{1}{|\tau_k||t_j(L_\beta+N)-\tau_k|} \frac{C^{N+1}F_j(L_\beta)}{\left|\prod_{n=0}^{N-1} t_j(L_\beta+n)\right| d_j^N} \times$$
$$\times \sum_{n=1}^{N+1}\left[\prod_{p=1}^{n-1}\tilde{D}_j(L_\beta+N-p)\right]\frac{\eta^{N+1-n}(\beta+N-q)!}{n!(\beta+N-q-n)!} \quad (106)$$

In the cases $|\tau_k - t_j(L_\beta+N)| > |\tau_k + t_j(L_\beta+N)|$ and $\tau_k = \tau_0 = 0$ expressions $|\tau_k||t_j(L_\beta+N)+\tau_k|$ and $|t_j(L_\beta+N)|d_j$ will be in the denominator of Eq. (106), respectively, instead of the expression $|\tau_k||t_j(L_\beta+N)-\tau_k|$. Considering that $[|\tau_k||t_j(L_\beta+N)\pm\tau_k|]^{-1} < \tilde{D}_j(L_\beta+N)[|t_j(L_\beta+N)|d_j]^{-1}$ and $\tilde{D}_j(l) \leq \hat{D}_j \cdot \ln(l+1)$, where $\hat{D}_j > 0$ is a constant, we have:

$$|A_{k,j}^{[L_\beta+N,q]}| \leq \frac{C^{N+1}F_j(L_\beta)\left[Max(1,\hat{D}_j)\cdot\ln(L_\beta+N+1)\right]^{N+1}\eta^N}{d_j^{N+1}\left|\prod_{n=0}^{N} t_j(L_\beta+n)\right|} \sum_{n=1}^{N+1}\frac{(\beta+N)!}{n!(\beta+N-n)!} \quad (107)$$

By using the relation: $\sum_{n=1}^{N+1}\binom{\beta+N}{n} \leq \sum_{n=0}^{\beta+N}\binom{\beta+N}{n} = 2^{\beta+N}$ [44], expression (107) can be simplified:

$$|A_{k,j}^{[L_\beta+N,q]}| \leq \frac{2^{\beta+N}C^{N+1}F_j(L_\beta)\left[Max(1,\hat{D}_j)\cdot\ln(L_\beta+N+1)\right]^{N+1}\eta^N}{d_j^{N+1}\left|\prod_{n=0}^{N} t_j(L_\beta+n)\right|}. \quad (108)$$

A similar expression is obtained in the case of $\tau_{k_{\beta+N}} = t_j(L_\beta+N)$. Thus, taking into account that $|t_j(l)| \geq 2l$ and setting $l' = L_\beta + N$, we find that the d'Alembert ratio test for the majorizing series (90) will be fulfilled for indexes:

$$l' > \gamma^2 C \cdot \eta \cdot Max(1,\hat{D}_j) \cdot \ln(l'+2) \cdot d_j^{-1}. \quad (109)$$

This implies that series (67) converges absolutely for any finite $\gamma$, which completes the proof.

In the case when the equality $\tau_{k_i} = t_j(l)$ is realized every $\Delta l$ iterations, the step in the sum over $n$ in Eq. (106) will be equal to $\Delta l$ and the degree of the constant $\eta$ will decrease $\Delta l$ times. Thus, in this case, we obtain a criterion similar to expression (109).

Let us now turn to the case of angles that are multiples of irrational factors of $\pi$. The complexity of the proof here lies in the fact that in this case we cannot define a finite $d_j$, since for any $\hat{d}_j(l)$ there can be an $l' > l$ such that $\hat{d}_j(l') < \hat{d}_j(l)$. As a result it is not so easy to choose the majorizing series of the form Eq. (74) with fixed $d_j$, and it is necessary to carefully investigate how the numbers $\tau_k$ behave in this case, which is beyond the scope of this paper. On the other hand, the case of angles that are multiples of irrational factors of $\pi$ is rather of theoretical interest. In practical calculations we use numbers with a finite number of digits in the mantissa, thus rounding irrational numbers to rational ones. In addition, in real physical applications the angles are specified with some tolerance, which cannot be less than a certain physical limit, since material objects consist of vibrating atoms and molecules, and the exact values of the angles are just an abstraction.

In the conclusion of this section, we note that in numerical calculations it is more convenient to make the replacement $\tilde{A}_{k,j}^{[l,q]} = d_j^q A_{k,j}^{[l,q]}$ so, instead of Eq. (88) we have:

$$R_{k,j}^{[l]}(r) = R_{k,j}^{[l-1]}(r) + \gamma^{2l} r^{t_j(l)} \sum_{q=0}^{Q} \frac{\tilde{A}_{k,j}^{[l,q]}}{q!} (\ln r)^q, \quad Q = \begin{cases} \beta_{l-1}+1, & \tau_k = t_j(l) \\ \beta_{l-1}, & otherwise \end{cases}, \tag{110}$$

where $t_j(l) \equiv \tau_j + 2l$, the coefficients $\tilde{A}_{k,j}^{[l,q]}$ are defined recurrently as nested series:

$$\tilde{A}_{k,j}^{[l,q]} = \begin{cases} \sum_{s=0}^{\infty} c_{k,s} \sum_{p=0}^{\beta_{l-1}-q} \frac{(-1)^{p+1}(p+1)}{t_j(l)^{p+2}} \tilde{A}_{s,j}^{[l-1,p+q]}, & \tau_k = 0 \\ \sum_{s=0}^{\infty} c_{k,s} \sum_{p=Max(1-q,0)}^{\beta_{l-1}+1-q} \frac{(-1)^{p+1}}{2^{p+1} t_j(l)^{p+1}} \tilde{A}_{s,j}^{[l-1,p+q-1]}, & \tau_k = t_j(l) \\ \sum_{s=0}^{\infty} c_{k,s} \sum_{p=0}^{\beta_{l-1}-q} \frac{[t_j(l)+\tau_k]^{p+1} - [t_j(l)-\tau_k]^{p+1}}{2\tau_k [\tau_k^2 - t_j(l)^2]^{p+1}} \tilde{A}_{s,j}^{[l-1,p+q]}, & otherwise \end{cases}, \tag{111}$$

and at the beginning of the iterative process $R_{k,j}^{[0]}(r) = \delta_{k,j} r^{\tau_j}$, $\tilde{A}_{k,j}^{[0,q]} = \delta_{k,j} \cdot \delta_{q,0}$, $\beta_0 = 0$. If in the process of calculating $\tilde{A}_{k,j}^{[l,q]}$ at least once the equality $\tau_k = t_j(l)$ occurs, then the number $\beta$ as a result of this iteration increases by 1:

$$\beta_l = \begin{cases} \beta_{l-1}+1, & if \ \exists k, \tau_k = t_j(l) \\ \beta_{l-1}, & otherwise \end{cases}. \tag{112}$$

Note also that in formulae (88)-(89) the number $\beta$ was used as $\beta_l$ in Eq. (111).

### 2.3 Calculation of radial functions R by direct solution of a system of differential equations

The first few radial functions can be also found by simpler but less accurate method, solving numerically the Cauchy problem for a system of differential equations. Let us represent the functions $R_{k,j}(r)$ like

$$R_{k,j}(r) = R_j^0(r) \cdot f_{k,j}(r), \tag{113}$$

From Eq. (66) it follows that $f_{k,j}(r)$ is written as

$$f_{k,j}(r) = \delta_{k,j} + \gamma^2 a_{k,j,1}(r) \cdot r^2 + \gamma^4 a_{k,j,2}(r) \cdot r^4 + \ldots, \tag{114}$$

where the functions $a_{k,l}(r)$ consist of the sum of the constants or the constants and the logarithms in an integer power. Therefore, since the power of the logarithms in any of the terms is less than that of $r$, the initial conditions for the function $f_{k,j}(r)$ and its derivative $f'_{k,j}(r)$ will be

$$f_{k,j}(0) = \delta_{k,j}, \quad f'_{k,j}(0) = 0. \tag{115}$$

Substituting Eq. (113) into Eq. (12) we obtain for each $j = 0..N_w - 1$ a system of $N_\Phi$ differential equations:

$$\left[ r\partial_r (r\partial_r) + 2\tau_j r\partial_r + (\tau_j^2 - \tau_k^2) \right] f_{k,j}(r) + r^2 \gamma^2 \sum_{l=0}^{N_\Phi - 1} c_{k,l} f_{l,j}(r) = 0, \quad k = 0..N_\Phi - 1. \tag{116}$$

## Appendix 3. The system of linear equations for the coefficients in the analytical modal method taking into consideration the edge singularities

In numerical calculations we restrict ourselves in expansion (19) to the first $M$ eigenfunctions, and in expansions (3) near each $w$-th of the total number of $W_l$ angular singularities, situated at the boundary with the coordinate $y_l$ between layers $l$-1 and $l$ in Fig. 2a, to the first $N_w$ functions. Substituting Eq. (19) for the field on the boundary between the layers $l$-1 and $l$ directly into Eq. (21), multiplying by complex conjugated test functions $\phi_n(x)$, which can be, for example, the first $M - \sum_{w=1..W_l} N_w/2$ terms of the complex Fourier series and integrating over segments $[x_{2j}(y_l)..x_{2j+1}(y_l)]$, $j=0,1,...$ between regions with angular singularities (in Fig. 2a) these are segments $[x_0..x_1]$, $[x_2..x_3]$ and $[x_4..x_5]$), we obtain a system of $2M - \sum_{w=1..W_l} N_w$ linear equations:

$$\mathbf{A}_{l-1,y_l}\left[\mathbf{e}_{l-1}\mathbf{c}^+_{l-1} + \tilde{\mathbf{c}}^-_{l-1}\right] = \mathbf{A}_{l,y_l}\left[\mathbf{c}^+_l + \mathbf{e}_l\tilde{\mathbf{c}}^-_l\right]$$
$$\mathbf{B}_{l-1,y_l}\left[\mathbf{e}_{l-1}\mathbf{c}^+_{l-1} - \tilde{\mathbf{c}}^-_{l-1}\right] = \mathbf{B}_{l,y_l}\left[\mathbf{c}^+_l - \mathbf{e}_l\tilde{\mathbf{c}}^-_l\right]$$
(117)

where the diagonal matrix $\mathbf{e}_l = \exp(ik_0\boldsymbol{\kappa}_l \Delta y_l)$, $\Delta y_l = y_{l+1} - y_l$, the matrices $\mathbf{A}_{l,y_l}$, $\mathbf{B}_{l,y_l}$ have the elements:

$$\left(\mathbf{A}_{l,y_l}\right)_{n,m} = \sum_j \int_{x_{2j}(y_l)}^{x_{2j+1}(y_l)} \phi_n^*(x)\psi_{l,m}(x)dx, \quad \left(\mathbf{B}_{l,y_l}\right)_{n,m} = \kappa_{l,m}\sum_j \varepsilon_{l,j}^{-1} \int_{x_{2j}(y_l)}^{x_{2j+1}(y_l)} \phi_n^*(x)\psi_{l,m}(x)dx, \quad (118)$$

and for the stability of the numerical solution of the linear system we made the substitution $c^-_{l,n} = \tilde{c}^-_{l,n} e_{l,n}$, see [31]. Physically this means that the expansion coefficients of upward and downward waves are defined at opposite sides of the layer. In layer $l$-1, at the boundary surrounding edge $w$ with center at point $(x_w, y_l)$ and radius $\rho_0$, conditions (21) are written:

$$\sum_{m=1}^{M}\left[c^+_{l-1,m}e_{l-1,m}e^+_{l-1,m}(\varphi) + \tilde{c}^-_{l-1,m}e^-_{l-1,m}(\varphi)\right]\psi_{l-1,m}(\varphi) = \sum_{j=0}^{N_w-1} h_j \sum_{k=0}^{N_w-1} \Phi_k(\varphi) R_{k,j}(1),$$

$$ik_0\sin\varphi\sum_{m=1}^{M}\kappa_{l-1,m}\left[c^+_{l-1,m}e_{l-1,m}e^+_{l-1,m}(\varphi) - \tilde{c}^-_{l-1,m}e^-_{l-1,m}(\varphi)\right]\psi_{l-1,m}(\varphi) + \quad , (119)$$

$$+\cos\varphi\sum_{m=1}^{M}\left[c^+_{l-1,m}e_{l-1,m}e^+_{l-1,m}(\varphi) + \tilde{c}^-_{l-1,m}e^-_{l-1,m}(\varphi)\right]\psi'_{l-1,m}(\varphi) = \sum_{j=0}^{N_w-1} h_j \sum_{k=0}^{N_w-1} \Phi_k(\varphi)\rho_0^{-1} R'_{k,j}(1)$$

where $e^\pm_{l,m}(\varphi) = \exp(\pm ik_0\kappa_{l,m}\rho_0\sin\varphi)$ and the signs were chosen taking into account the fact that the $y$ axis is directed downward, the functions $\Phi_k(\varphi)$, $R_{k,j}(1)$ and the coefficients $h_j$ refer to the $w$-th edge, for brevity we do not denote this as an additional index, we also used the abbreviated notation $\psi_{l,m}(\varphi) = \psi_{l,m}[x(\rho_0,\varphi)] = \psi_{l,m}(x_w + \rho_0\cos\varphi)$, $\psi'_{l,m}(\varphi) = \partial_x\psi_{l,m}[x(\rho_0,\varphi)]$. When writing Eq. (119) we took into account that $\partial_\rho R_{k,j}(\rho/\rho_0)\big|_{\rho=\rho_0} = \rho_0^{-1} R'_{k,j}(1)$ and truncate $R_{k,j}$ to a square matrix by setting $N_\Phi = N_w$. Multiplying Eq. (119), and similar equations written for a part of the boundary around an edge in the layer $l$, by complex conjugate adjoint eigenfunctions $\Phi_n^{+*}(\varphi)$ with weight $\varepsilon^{-1}(\varphi)$ and by integrating over the angle, taking into account orthogonality, we obtain a system of $2N_w$ equations:

$$\mathbf{A}^+_{l-1}\mathbf{e}_{l-1}\mathbf{c}^+_{l-1} + \mathbf{A}^-_{l-1}\mathbf{c}^-_{l-1} + \mathbf{A}^+_l\mathbf{c}^+_l + \mathbf{A}^-_l\mathbf{e}_l\tilde{\mathbf{c}}^-_l = \mathbf{R}\cdot\mathbf{h}$$
$$\mathbf{B}^+_{l-1}\mathbf{e}_{l-1}\mathbf{c}^+_{l-1,m} - \mathbf{B}^-_{l-1}\mathbf{c}^-_{l-1} + \mathbf{B}^+_l\mathbf{c}^+_{l,m} - \mathbf{B}^-_l\mathbf{e}_l\tilde{\mathbf{c}}^-_l = \rho_0^{-1}\mathbf{R}'\cdot\mathbf{h}$$
(120)

where the matrix elements $A^\pm_{l-1,n,m}$, $B^\pm_{l-1,n,m}$, $A^\pm_{l,n,m}$, and $B^\pm_{l,n,m}$ are:

$$A^\pm_{l-1,n,m} = \int_\pi^{2\pi} \Phi_n^{+*}(\varphi)\varepsilon^{-1}_{l-1}(\varphi) e^\pm_{l-1,m}(\varphi)\psi_{l-1,m}(\varphi)d\varphi$$

$$B^\pm_{l-1,n,m} = \int_\pi^{2\pi} \Phi_n^{+*}(\varphi)\varepsilon^{-1}_{l-1}(\varphi)\left[ik_0\sin(\varphi)\kappa_{l-1,m}\psi_{l-1,m}(\varphi) \pm \cos(\varphi)\psi'_{l-1,m}(\varphi)\right]e^\pm_{l-1,m}(\varphi)d\varphi$$

$$A^\pm_{l,n,m} = \int_0^\pi \Phi_n^{+*}(\varphi)\varepsilon^{-1}_l(\varphi) e^\pm_{l,m}(\varphi)\psi_{l,m}(\varphi)d\varphi$$

$$B^\pm_{l,n,m} = \int_0^\pi \Phi_n^{+*}(\varphi)\varepsilon^{-1}_l(\varphi)\left[ik_0\sin(\varphi)\kappa_{l,m}\psi_{l,m}(\varphi) \pm \cos\varphi\cdot\psi'_{l,m}(\varphi)\right]e^\pm_{l,m}(\varphi)d\varphi$$
(121)

By expressing the coefficients **h** of the expansion inside the region containing the edge in terms of the coefficients **c**, we obtain $N_w$ equations:

$$\left(\mathbf{A}_{l-1}^+ - \mathbf{D}^{-1}\mathbf{B}_{l-1}^+\right)\mathbf{e}_{l-1}\mathbf{c}_{l-1}^+ + \left(\mathbf{A}_{l-1}^- + \mathbf{D}^{-1}\mathbf{B}_{l-1}^-\right)\tilde{\mathbf{c}}_{l-1}^- + \left(\mathbf{A}_l^+ - \mathbf{D}^{-1}\mathbf{B}_l^+\right)\mathbf{c}_l^+ + \left(\mathbf{A}_l^- + \mathbf{D}^{-1}\mathbf{B}_l^-\right)\mathbf{e}_l\tilde{\mathbf{c}}_l^- = 0, \quad (122)$$

where the matrix $\mathbf{D} = \rho_0^{-1}\mathbf{R}'(1)\mathbf{R}^{-1}(1)$ is the Dirichlet-to-Neumann map operator matrix in the representation of the functions $\Phi_i(\varphi)$, see Appendix 4. After the same is done for the remaining edges between the layers $l$-1 and $l$ we obtain together with Eq. (117) $2M$ equations in total. In the same way the equations for the coefficients of all layers are written. The right-hand side is obtained by substituting the known coefficients $c_{0,m}^+ = \delta(0,m)$ for the incident monochrome wave, where $\delta(0,m)$ is the Kronecker symbol, and the zero value for the coefficients $\mathbf{c}_L^- \equiv 0$ of the last layer in Eq. (117) and Eq. (122), and transferring the result to the right-hand side.

## Appendix 4. Calculation of the Dirichlet-to-Neumann map operator matrix and the solution of the Dirichlet problem for a domain containing an edge

To obtain the DtN map operator matrix $\mathbf{D}_S$ of a region containing an edge and restricted by boundary $\rho = \rho_0$, it is convenient to use the functions $G_j(\rho/\rho_0 = 1,\varphi)$, $j = 0..N_w - 1$, Eq. (9), as a set of linearly independent boundary conditions. Matrices of these linearly independent vectors of boundary conditions and the corresponding normal derivatives at $N_w$ points $\varphi_i$, for which we can take the points of zeros of orthogonal polynomials, for example, Legendre or Chebyshev in each sector of uniform permittivity, will have the elements:

$$F_{i,j} \equiv G_j(1,\varphi_i) = \sum_{k=0}^{N_w-1} \Phi_k(\varphi_i)R_{k,j}(1), \quad \partial_\mathbf{n}F_{i,j} \equiv \partial_\mathbf{n}G_j(1,\varphi_i) = \sum_{k=0}^{N_w-1} \Phi_k(\varphi_i)\rho_0^{-1}R'_{k,j}(1). \quad (123)$$

Whence, using Eq. (23) we obtain the DtN map operator matrix $\mathbf{D}_S$ acting on the functions defined at the collocation points $\varphi_i$.

It is also easy to obtain the DtN map operator matrix $\mathbf{D}_S^{(\Phi)}$ in the representation of functions $\Phi_i(\varphi)$, when this matrix acts on the expansion coefficients $c_i$ of the conditions $f_\Gamma(\varphi)$ at the boundary $\rho = \rho_0$ in the functions $\Phi_i(\varphi)$, and as a result we obtain the expansion coefficients $c'_i$ of the corresponding normal derivative $\partial_\mathbf{n}f_\Gamma(\varphi)$ in these functions:

$$f_\Gamma(\varphi) = \sum_{i=0}^{N_w-1} \Phi_i(\varphi)c_i, \quad \partial_\mathbf{n}f_\Gamma(\varphi) = \sum_{j=0}^{N_w-1} N_j(\varphi)c_j = \sum_{j=0}^{N_w-1} \Phi_j(\varphi)c'_j, \quad (124)$$

where $N_i(\varphi)$ are normal derivatives of the solutions of Eq. (1) corresponding to the boundary conditions $\Phi_i(\varphi)$. Multiplying the last equality in Eq. (124) by the complex conjugate adjoint eigenfunctions $\Phi_i^{+*}(\varphi)$ and integrating over $\varphi$ with weight $\varepsilon^{-1}(\varphi)$, taking into account orthogonality, we find the expression for the coefficients of the matrix $\mathbf{D}_S^{(\Phi)}$:

$$c'_i = \sum_{j=0}^{N_w-1} \left[\int_0^{2\pi} \Phi_i^{+*}(\varphi)\varepsilon^{-1}(\varphi)N_j(\varphi)d\varphi\right]c_j = \sum_{j=0}^{N_w-1} D_{S,i,j}^{(\Phi)}c_j. \quad (125)$$

To find the normal derivatives $N_i(\varphi)$, we formally decompose $\Phi_i(\varphi)$ in functions $G_j(1,\varphi)$ (9):

$$\Phi_i(\varphi) = \sum_{j=0}^{N_w-1} h_{j,i}G_j(1,\varphi) = \sum_{j=0}^{N_w-1} h_{j,i}\sum_{k=0}^{N_w-1} \Phi_k(\varphi)R_{k,j}(1) \quad (126)$$

Multiplying Eq. (126) by $\Phi_l^{+*}(\varphi)$ and integrating over $\varphi$ with weight $\varepsilon^{-1}(\varphi)$, gives us the coefficient matrix: $\mathbf{h} = \mathbf{R}^{-1}(1)$. Using the found coefficients $h_{j,i}$, we can write the expression for the normal derivatives $N_i(\varphi)$ corresponding to $\Phi_i(\varphi)$ on the boundary $\rho = \rho_0$:

$$N_i(\varphi) = \sum_{j=0}^{N_w-1} h_{j,i} \partial_\mathbf{n} G_j(1,\varphi) = \sum_{j=0}^{N_w-1} R_{j,i}^{(-1)}(1) \sum_{k=0}^{N_w-1} \Phi_k(\varphi) \rho_0^{-1} R'_{k,j}(1). \qquad (127)$$

Thus, substituting $N_i(\varphi)$ into Eq. (125) we obtain the DtN map operator matrix $\mathbf{D}_S^{(\Phi)}$ in the representation of functions $\Phi_i(\varphi)$:

$$\mathbf{D}_S^{(\Phi)} = \rho_0^{-1} \mathbf{R}'(1) \mathbf{R}^{-1}(1). \qquad (128)$$

It is more convenient to work in the representation of coefficients, since it does not require equality of the number of collocation points at adjacent boundaries. In this work we used this very representation, connecting firstly all the elements around a region containing an edge into the single element with the DtN map operator matrix $\mathbf{D}$ by means of (25) and after that changing representation on the boundary of this region from the values at collocation points $\varphi_i$ to the values of expansion coefficients in the eigenfunctions $\{\Phi_k(\varphi)\}$. Then we join it to the region containing an edge characterized by DtN map operator matrix $\mathbf{D}_S^{(\Phi)}$.

To change the representation in DtN map operator matrix $\mathbf{D}$ at part $\Gamma_S$ of the boundary to the basis functions $\{\Phi_k(\varphi)\}$ representation without changing the representation on the other part $\Gamma$ of the boundary, let us to represent an arbitrary boundary condition $f_{\Gamma_S}(\varphi_i)$ and its corresponding normal derivative $\partial_\mathbf{n} f_{\Gamma_S}(\varphi_i)$ in the basis functions $\{\Phi_k(\varphi_i)\}$ at Gauss or Gauss-Lobatto quadrature points $\varphi_i$. In matrix form this can be written:

$$\mathbf{f}_{\Gamma_S} = \mathbf{\Phi} \cdot \mathbf{c}_{\Gamma_S}, \quad \partial_\mathbf{n} \mathbf{f}_{\Gamma_S} = \mathbf{\Phi} \cdot \mathbf{c}'_{\Gamma_S}, \qquad (129)$$

where the matrix $\mathbf{\Phi}$ consists of the columns of the basis functions $\Phi_k(\varphi_i)$, and $\mathbf{c}_{\Gamma_S}$, $\mathbf{c}'_{\Gamma_S}$ are the coefficients columns. The action of the DtN map operator matrix $\mathbf{D}$ on the function at the border, which includes the parts $\Gamma_S$ and $\Gamma$, will be

$$\begin{pmatrix} \mathbf{D}_{1,1} & \mathbf{D}_{1,2} \\ \mathbf{D}_{2,1} & \mathbf{D}_{2,2} \end{pmatrix} \begin{pmatrix} \mathbf{f}_{\Gamma_S} \\ \mathbf{f}_\Gamma \end{pmatrix} = \begin{pmatrix} \partial_\mathbf{n} \mathbf{f}_{\Gamma_S} \\ \partial_\mathbf{n} \mathbf{f}_\Gamma \end{pmatrix} \Leftrightarrow \begin{pmatrix} \mathbf{D}_{1,1} & \mathbf{D}_{1,2} \\ \mathbf{D}_{2,1} & \mathbf{D}_{2,2} \end{pmatrix} \begin{pmatrix} \mathbf{\Phi} & 0 \\ 0 & \mathbf{I} \end{pmatrix} \begin{pmatrix} \mathbf{c}_{\Gamma_S} \\ \mathbf{f}_\Gamma \end{pmatrix} = \begin{pmatrix} \mathbf{\Phi} & 0 \\ 0 & \mathbf{I} \end{pmatrix} \begin{pmatrix} \mathbf{c}'_{\Gamma_S} \\ \partial_\mathbf{n} \mathbf{f}_\Gamma \end{pmatrix}, \qquad (130)$$

where $\mathbf{f}_\Gamma$ and $\partial_\mathbf{n} \mathbf{f}_\Gamma$ are the columns of the function values and the corresponding normal derivative on the boundary part $\Gamma$, $\mathbf{I}$ is the identity matrix. Multiplying Eq. (130) by the transposed matrix of complex conjugate adjoint eigenfunctions $\tilde{\mathbf{\Phi}}$ with the diagonal matrix of the corresponding integration weights $\mathbf{w}$, we obtain the DtN map operator matrix $\mathbf{D}$ in the mixed representation:

$$\begin{pmatrix} \tilde{\mathbf{\Phi}}\mathbf{w} & 0 \\ 0 & \mathbf{I} \end{pmatrix} \begin{pmatrix} \mathbf{\Phi} & 0 \\ 0 & \mathbf{I} \end{pmatrix} = \begin{pmatrix} \mathbf{I} & 0 \\ 0 & \mathbf{I} \end{pmatrix} \Rightarrow \begin{pmatrix} \tilde{\mathbf{\Phi}}\mathbf{w}\mathbf{D}_{1,1}\mathbf{\Phi} & \tilde{\mathbf{\Phi}}\mathbf{w}\mathbf{D}_{1,2} \\ \mathbf{D}_{2,1}\mathbf{\Phi} & \mathbf{D}_{2,2} \end{pmatrix} \begin{pmatrix} \mathbf{c}_{\Gamma_S} \\ \mathbf{f}_\Gamma \end{pmatrix} = \begin{pmatrix} \mathbf{c}'_{\Gamma_S} \\ \partial_\mathbf{n} \mathbf{f}_\Gamma \end{pmatrix}, \qquad (131)$$

If the matrix $\mathbf{\Phi}$ is square, then for $\tilde{\mathbf{\Phi}}\mathbf{w}$ we can simply take the inverse matrix $\mathbf{\Phi}^{-1}$. If the number of quadrature points $\varphi_i$ is greater than the number of basis functions $\{\Phi_k(\varphi_i)\}$, then equalities in (131) will be fulfilled with spectral accuracy inherent to Gauss quadratures.

After the conditions $b(\varphi_i)$ in the points $\varphi_i$ of the boundary $\rho = \rho_0$ of the region containing the edge were received, the coefficients $h_j$ of the expansion of the field (15) in the functions $G_j(\rho/\rho_0 = 1, \varphi_i)$ can be found from the equations:

$$b(\varphi_i) = \sum_{j=0}^{N_w-1} h_j G_j(\rho/\rho_0 = 1, \varphi_i) \Leftrightarrow \mathbf{G} \cdot \mathbf{h} = \mathbf{b}. \qquad (132)$$

If the boundary conditions are represented in the form of expansion coefficients in the functions $\Phi_i(\varphi)$:

$$b(\varphi) = \sum_{i=0}^{N_w-1} \Phi_i(\varphi) c_i = \sum_{j=0}^{N_w-1} h_j G_j(\rho/\rho_0 = 1, \varphi_i) = \sum_{j=0}^{N_w-1} h_j \sum_{k=0}^{N_w-1} \Phi_k(\varphi) R_{k,j}(\rho/\rho_0 = 1) \quad (133)$$

then, multiplying on the left side by $\Phi_l^{+*}(\varphi)$, and integrating with the weight $\varepsilon^{-1}(\varphi)$, we obtain a similar equation for the coefficients $h_j$ in the matrix notation:

$$c_i = \sum_{j=0}^{N_w-1} R_{i,j}(\rho/\rho_0 = 1) h_j \Leftrightarrow \mathbf{R} \cdot \mathbf{h} = \mathbf{c}. \quad (134)$$

## Acknowledgments

The author is grateful to Prof. Lifeng Li for his valuable comments on the paper.